\documentclass[12pt]{article}

\usepackage{amssymb}
\usepackage{amsmath}
\usepackage{amscd}
\usepackage{latexsym}
\usepackage{graphicx}
\usepackage{cite}

\newcommand{\be}{\begin{equation}}
\newcommand{\ee}{\end{equation}}
\newcommand{\Dlt}{\Delta}
\newcommand{\dlt}{\delta}
\newcommand{\prt}{\partial}
\newcommand{\br}{{\bf r}}
\newcommand{\bbe}{{\bf e}}
\newcommand{\bd}{{\bf d}}
\newcommand{\bj}{{\bf j}}
\newcommand{\bn}{{\bf n}}
\newcommand{\bk}{{\bf k}}
\newcommand{\bP}{{\bf P}}
\newcommand{\bA}{{\bf A}}
\newcommand{\bH}{{\bf H}}
\newcommand{\bJ}{{\bf J}}
\newcommand{\bE}{{\bf E}}
\newcommand{\bS}{{\bf S}}
\newcommand{\bM}{{\bf M}}
\newcommand{\bB}{{\bf B}}
\newcommand{\bt}{\beta}
\newcommand{\vp}{\varphi}
\newcommand{\vth}{\vartheta}
\newcommand{\ep}{\varepsilon}
\newcommand{\al}{\alpha}
\newcommand{\ra}{\rightarrow}

\newcommand{\gm}{\gamma}
\newcommand{\om}{\omega}
\newcommand{\Om}{\Omega}
\newcommand{\Gm}{\Gamma}
\newcommand{\dgr}{\dagger}
\newcommand{\lbd}{\lambda}

\newcommand{\cD}{{\cal D}}

\newcommand{\cH}{{\cal H}}

\newcommand{\rgl}{\rangle}
\newcommand{\lgl}{\langle}

\begin{document}

\begin{center}

{\Large{\bf Coherent dynamics of radiating atomic systems 
in pseudospin representation} \\ [5mm]

V.I. Yukalov } \\ [3mm]

{\it Bogolubov Laboratory of Theoretical Physics, \\
Joint Institute for Nuclear Research, Dubna 141980, Russia}

\vskip 2mm

E-mail: yukalov@theor.jinr.ru

\end{center}

\vskip 0.5cm

\begin{abstract}
The aim of this review is twofold. First, a general approach is presented 
allowing for a unified description of dynamics in radiating systems of 
different nature. Both atomic systems as well as spin assemblies can be 
treated in the frame of the same mathematical method based on pseudospin 
(or spin) representation of evolution equations. The approach is applicable 
to all stages of radiation dynamics, including the most difficult initial 
quantum stage, where coherence is not yet developed. This makes it possible 
to study the process of coherent self-organization from the chaotic quantum 
stage. Second, the approach is illustrated by applying it for the description 
of several coherent phenomena. Different types of superradiance are 
characterized: pure superradiance, triggered superradiance, pulsing and 
punctuated superradiance. The theory is presented of such interesting effects 
as triggering dipolar waves, turbulent photon filamentation, collective 
liberation of light, pseudospin atomic squeezing, and operator entanglement 
production. 
\end{abstract}

\newpage

{\bf Contents}

\vskip 2mm
1. Introduction

\vskip 2mm
2. Equations in pseudospin representation

\vskip 2mm
3. Self-action of a radiating atom

\vskip 2mm
4. Stochastic mean-field quantization

\vskip 2mm
5. Scale separation approach

\vskip 2mm
6. Magnetic dipole transitions

\vskip 2mm
7. Coherent and incoherent radiation

\vskip 2mm
8. Influence of external fields

\vskip 2mm
9. Triggering dipolar waves

\vskip 2mm
10. Transverse mode expansion

\vskip 2mm
11. Emergence of coherence from chaos

\vskip 2mm
12. Pulsing and punctuated superradiance

\vskip 2mm
13. Turbulent photon filamentation

\vskip 2mm
14. Collective liberation of light

\vskip 2mm
15. Pseudospin atomic squeezing

\vskip 2mm
16. Operator entanglement production

\vskip 2mm
17. Conclusion

\newpage

\section{Introduction}

Coherent radiation from atomic and molecular systems is well studied,
being the basis of laser radiation. The theory of this radiation, occurring 
in optical or infrared diapason, is thoroughly expounded in many books
(see, e.g. \cite{Allen_1,Andreev_2,Mandel_3,Trifonov_4,Rubtsova_5}). Another 
type of systems whose collective radiation has recently attracted much 
attention are spin assemblies \cite{Yukalov_6}. Atomic and spin systems, 
being rather different in nature, are usually described by different types 
of equations. Dynamical physical processes in these systems are really quite 
different. However, it is possible to develop a mathematical approach that 
would allow for a similar description of both these system types. Such an 
approach is presented in this review, where we concentrate on the dynamics
of atomic systems, illustrating the approach by the description of several 
coherent phenomena.    

A problem of great interest is the self-organization of coherence from  
initially chaotic quantum fluctuations. The developed approach gives the 
possibility of analyzing such a self-organization in detail. Briefly speaking,
the basic mathematical points of the approach are: pseudopsin representation
of evolution equations, stochastic quantization, scale separation, transverse 
mode expansion, and probabilistic pattern selection. 

As illustrations of the approach, several nontrivial phenomena will be 
treated, such as triggering dipolar waves, different kinds of superradiance
(pure superradiance, triggered superradiance, pulsing and punctuated 
superradiance), turbulent photon filamentation, collective liberation of light, 
pseudospin atomic squeezing, and operator entanglement production. 

The main point, emphasized throughout the paper, is the generality of the used
mathematical techniques that can be applied to any kind of evolution equations
describing radiating systems. It is evident that the equations of motion for 
spin systems are written for the spin degrees of freedom. The pseudospin 
representation for atomic systems makes it straightforward to employ the same 
mathematical techniques for spin as well as for atomic radiating systems.   

This review is planned for the Special Issue devoted to the memory of 
Igor V. Yevseyev. Igor was my friend for many years and I had a pleasure of 
discussing with him various scientific and non-scientific problems. My memory
about Igor is interconnected with those problems we discussed. Exactly this 
interconnection dictated the choice of the material for this review, in which 
I included the material I had discussed with Igor.

\section{Equations in pseudospin representation}  

In order to reduce the equations for radiating atomic systems to the form 
that would also be convenient for treating spin systems, it is natural to 
resort to pseudospin representation and eliminate the field degrees of freedom
\cite{Yukalov_7,Yukalov_8}. In addition, aiming at accurately characterizing 
the self-organization of coherence from chaos, it is necessary to start with 
a microscopic picture. 

Let us consider $N$ atoms (or molecules) resonantly interacting with 
electromagnetic field. The general form of the Hamiltonian is
\be
\label{1}
 \hat H =  \hat H_a + \hat H_f + \hat H_{af} + \hat H_{mf} \; .
\ee
The first term 
\be
\label{2}
\hat H_a = \sum_{j=1}^N \om_0 \left ( \frac{1}{2} + S_j^z \right )
\ee
corresponds to resonant atoms with the transition frequency $\omega_0$; 
the pseudospin operator $S_j^z$ characterizes interlevel electronic transitions 
of a $j$th atom. Considering here two-level atoms, we deal with the pseudospin 
operators of spin one-half. Atoms with a larger number of resonant levels would 
require the use of higher-order pseudospin operators. The operators are called 
pseudospin because they satisfy the spin algebra, but do not describe real spins,
characterizing instead interlevel transitions.  

Here and in what follows, the system of units is employed where the Planck 
constant is set to unity.

The second term in equation (1) defines the energy of radiated electromagnetic 
field
\be
\label{3}
\hat H_f = \frac{1}{8\pi} \int \left ( \bE^2 + \bH^2 \right ) \; d\br \;  ,
\ee
where ${\bf E}$ is electric field and ${\bf H}$ is magnetic field represented 
through vector potential ${\bf A}$,
\be
\label{4}
 \bH = \nabla \times \bA \;  .
\ee
The vector potential is assumed to satisfy the Coulomb gauge calibration
\be
\label{5}
\nabla \cdot \bA  = 0 \;  .
\ee

The third term describes the atom-field interaction
\be
\label{6}
 \hat H_{af} = -\sum_{j=1}^N  \left ( \frac{1}{c}\; \bJ_j \cdot \bA_j + 
\bP_j \cdot \bE_{0j} \right ) \; ,
\ee
where dipolar transitions are assumed, $\bE_{0j}$ is an external electric 
field, the current operator is
\be
\label{7}
\bJ_j = - i\om_0 \left ( \bd^* S_j^- - \bd S_j^+ \right )
\ee
and the polarization operator is
\be
\label{8}
\bP_j =  \bd^* S_j^- + \bd S_j^+ \; ,
\ee
with the ladder pseudospin operators
$$
 \bS_j \equiv S_j^x \pm i S_j^y \;  ,
$$
and ${\bf d}$ being a transition dipole. The notation
$$
\bA_j \equiv \bA(\br_j,t) \; , \qquad \bE_{0j} \equiv \bE_0(\br_j,t)
$$
is used. 

The last term in equation (1) describes the interaction of the radiated 
field with the matter surrounding atoms. This term is absent when atoms are 
in vacuum. But if atoms are immersed into some kind of matter, the 
interaction term is
\be
\label{9}
\hat H_{mf} = - \; \frac{1}{c} \int \bj_{mat} \cdot \bA \;  d\br \;  ,
\ee
where ${\bf j}_{mat}$ is the density current in the matter. 

The evolution equations are prescribed by the Heisenberg equations of motion,
with the corresponding commutation relations. The electromagnetic operators
satisfy the relations
$$
\left [ E^\al(\br,t) , \; A^\bt(\br',t) \right ] = 
4\pi i c \dlt_{\al\bt}(\br-\br') \; ,
\qquad
\left [ A^\al(\br,t) , \; H^\bt(\br',t) \right ] = 0 \; ,
$$
$$
 \left [ E^\al(\br,t) , \; H^\bt(\br',t) \right ] = -4\pi i c \sum_\gm
\ep_{\al\bt\gm} \; \frac{\prt}{\prt r^\gm}\; \dlt(\br-\br') \; ,
$$
in which $\varepsilon_{ijk}$ is the unitary antisymmetric tensor, $c$ is light 
velocity, and the transverse delta function is defined as
$$
\dlt_{\al\bt}(\br) = \frac{2}{3} \; \dlt_{\al\bt} \dlt(\br) - \;
\frac{1}{4\pi} \; D_{\al\bt}(\br) \;   ,
$$
with the dipolar tensor
$$
D_{\al\bt}(\br) \equiv \frac{1}{r^3} \left ( \dlt_{\al\bt} - 3n^\al n^\bt
\right ) \;  ,
$$
where
$$
r \equiv | \br | \; , \qquad \bn \equiv \frac{\br}{r} \;   .
$$
The pseudospin operators obey the spin algebra
$$
 \left [ S_j^- , \; S_i^+ \right ] = - 2 \dlt_{ij} S_j^z \; , \qquad
 \left [ S_j^- , \; S_i^z \right ] =  \dlt_{ij} S_j^- \; , \qquad
\left [ S_j^+ , \; S_i^z \right ] = - \dlt_{ij} S_j^+ \; .
$$
 
The Heisenberg equations for the field variables yield the Maxwell equations
\be
\label{10}
  \frac{1}{c} \; \frac{\prt \bE}{\prt t} = 
\nabla \times \bH - \; \frac{4\pi}{c} \; \bj \; , \qquad   
\frac{1}{c} \; \frac{\prt \bA}{\prt t} = - \bE \; ,
\ee
from where, with the Coulomb calibration (5), the wave equation follows:
\be
\label{11}
 \left (  \nabla^2 - \; \frac{1}{c^2} \; \frac{\prt^2}{\prt t^2}\right )\bA =
- \; \frac{4\pi}{c}\; \bj \;  ,
\ee
with the density of current
\be
\label{12}
 j^\al(\br,t) = \sum_\bt \left [ \sum_{i=1}^N \dlt_{\al\bt}(\br-\br_i) J_i^\bt(t)
+ \int \dlt_{\al\bt}(\br-\br') j_{mat}^\bt(\br',t) \; d\br ' \right ] \;  .
\ee
The solution to equation (11) reads
\be
\label{13}
 \bA(\br,t) = \bA_{vac}(\br,t) + \frac{1}{c} \int \bj \left (\br', t - \;
\frac{|\br-\br'|}{c} \right ) \frac{d\br'}{|\br-\br'|} \;  ,
\ee
where ${\bf A}_{vac}$ is the vacuum vector potential that is a solution 
to the equation
\be
\label{14}
 \left ( \nabla^2 - \; \frac{1}{c^2} \; 
\frac{\prt^2}{\prt t^2}\right )\bA_{vac} = 0 \; .
\ee

The Heisenberg equations for the pseudospin variables give
$$
\frac{dS_j^-}{dt} = - i\om_0 S_j^- + 2S_j^z \left ( k_0 \bd \cdot \bA_j
- i \bd \cdot \bE_{0j} \right ) \; ,
$$
\be
\label{15}
 \frac{dS_j^z}{dt} = -  S_j^+  \left ( k_0 \bd \cdot \bA_j
- i \bd \cdot \bE_{0j} \right ) - 
S_j^- \left ( k_0 \bd^* \cdot \bA_j + 
i \bd^* \cdot \bE_{0j} \right ) \;  ,
\ee
where $k_0 = \omega_0/c$.  

The vector potential (13), with the current density (12), can be written, 
excluding self-action, as
\be
\label{16}
 \bA = \bA_{vac} + \bA_{rad} + \bA_{dip} + \bA_{mat} \;  .
\ee
The first term here is due to radiating atoms,
\be
\label{17}
 \bA_{rad}(\br_i,t) = \sum_{j(\neq i)}^N \frac{2}{3cr_{ij} } \;
\bJ_j \left ( t - \; \frac{r_{ij}}{c} \right ) \;  ,
\ee
where
$$
r_{ij} \equiv | \br_{ij} | \; , \qquad 
\br_{ij} \equiv \br_i - \br_j \;   .
$$
Using the form of current (7), we find 
\be
\label{18}
 \bA_{rad}(\br_i,t) = -i \sum_{j(\neq i)}^N \frac{2k_0}{3r_{ij} } 
\left [ \bd^* S_j^- \left ( t - \; \frac{r_{ij}}{c} \right ) -
\bd S_j^+ \left ( t - \; \frac{r_{ij}}{c} \right )  \right ] \;  .
\ee
The second term in equation (16) is caused by induced atomic dipoles,
\be
\label{19}
 A^\al_{dip}(\br,t) = - \sum_{j=1}^N \sum_\bt \int 
\frac{D_{\al\bt}(\br'-\br_j)}{4\pi c|\br - \br'|} \; 
J_j^\bt \left ( t - \; \frac{|\br-\br'|}{c} \right ) \; d\br' \; .
\ee
And the last term 
\be
\label{20}
 A_{mat}^\al(\br,t) =  \sum_\bt \int
\frac{\dlt_{\al\bt}(\br'-\br'')}{c|\br - \br'|} \;
j_{mat}^\bt \left (r'', t - \; \frac{|\br-\br'|}{c} \right ) \; 
d\br' d\br''
\ee
corresponds to the vector potential created by the current in the matter. 

Notice that, due to the dependence on the variable $t - r/c$, we have
\be
\label{21}
\left ( \frac{1}{c} \frac{\prt}{\prt t} + \frac{\prt}{\prt r} \right )
S_j^\pm  \left ( t - \; \frac{r}{c} \right ) = 0 \;  .
\ee
 
We assume that electromagnetic fields do not strongly disturb atomic level
structure, so that  
\be
\label{22}
  \frac{|\bd\cdot \bE|}{\om_0} \ll 1 \; , \qquad 
 \frac{|\bd\cdot \bE_0|}{\om_0} \ll 1 \; .
\ee
Then the retardation effects can be described in the Born approximation. 
Under condition (22), from equations (15) it follows
\be
\label{23}
\frac{\prt}{\prt r}\; S_j^\pm  \left ( t - \; \frac{r}{c} \right )
= \mp ik_0 S_j^\pm  \left ( t - \; \frac{r}{c} \right )  .
\ee
Setting the retardation condition
\be
\label{24}
 S_j^\al(t) = 0 \qquad ( t < 0 ) \;  ,
\ee
we finally get in the Born approximation
\be
\label{25}
 S_j^- \left ( t - \; \frac{r}{c} \right ) = 
S_j^-(t) \Theta(ct-r) e^{ik_0 r} \; , \qquad
S_j^z \left ( t - \; \frac{r}{c} \right ) = 
S_j^z(t) \Theta(ct-r) \; .
\ee

\section{Self-action of a radiating atom}

Strictly speaking, substituting the current density (12) into the integral
in equation (13), one meets the terms corresponding to the atomic self-action, 
which can be treated as follows. The vector potential generated by a single 
atom is
\be
\label{26}
A^\al_s(\br,t) = \frac{1}{c} \sum_\bt  \; \int
\frac{\dlt_{\al\bt}(\br')}{|\br-\br'|} \;
J^\bt \left ( t -\; \frac{|\br-\br'|}{c}\right ) \; d\br' \; ,
\ee
with the current
$$
\bJ\left ( t - \; \frac{r}{c}\right ) = i\om_0\left [
\bd S^+(t)\; e^{-ik_0r} -\bd^*\; S^-(t)\; e^{ik_0r}\right ]\;
\Theta(ct-r) \; ,
$$
where $S^\al(t)\equiv S^\al(0,t)$. At small distance, such that
$k_0r\ll 1$, one may write $e^{ik_0r}\simeq 1+ik_0r$. Substituting the
transverse $\dlt$-function into the vector potential (26), we keep
in mind that averaging the dipolar tensor over spherical angles
gives
$$
\int D_{\al\bt}(\br) \; d\Omega(\br) = 0 \; .
$$
Then, for $k_0r\ll 1$, the vector potential (26) becomes
$$
\bA_s(\br,t) \simeq \frac{2}{3}\; k_0^2\left [ \bd S^+(t) +
\bd^* S^-(t)\right ] + i\; \frac{2k_0}{3r}\; \left [
\bd S^+(t) - \bd^* S^-(t)\right ] \; .
$$
To avoid the divergence in the term $1/r$, one can average it between the
electron wavelength $\lbd_e=2\pi\hbar/mc$, with $m$ being the electron
mass, and the radiation wavelength $\lbd_0=2\pi/k_0$. Taking into account
that $\lbd_e\ll\lbd_0$, we have
$$
\frac{1}{\lbd_0-\lbd_e} \; \int_{\lbd_e}^{\lbd_0} \frac{dr}{r} =
\frac{k_0}{2\pi}\; \ln\left ( \frac{mc^2}{\hbar\om_0} \right ) \; .
$$
Then for the self-acting vector potential, we get
\be
\label{27}
\bA_s(0,t) = \frac{2}{3}\; k_0^2 \left [ \bd S^+(t) + \bd^* S^-(t)
\right ] + \frac{ik_0}{3\pi}\; \ln\left ( \frac{mc^2}{\hbar\om_0}\right )
\left [ \bd S^+(t) - \bd^* S^-(t) \right ] \; .
\ee
Substituting this into the evolution equations for the case of a single atom, 
we employ the properties of spin one-half operators. Then we come to the 
equations for a single atom
\be
\label{28}
\frac{dS^-}{dt} = - i ( \om_0 -\dlt_L - i\gm_0) S^- +
\frac{\bd^2}{|\bd|^2}\; (\gm_0 + i\dlt_L) S^+ \; , \qquad
\frac{dS^z}{dt} = -2\gm_0\left ( \frac{1}{2} + S^z \right ) \; ,
\ee
in which the notations for the natural width
\be
\label{29}
\gm_0 \equiv \frac{2}{3}\; |\bd|^2 k_0^3
\ee
and the Lamb shift
\be
\label{30}
\dlt_L \equiv \frac{\gm_0}{2\pi} \; \ln \left (
\frac{mc^2}{\hbar\om_0}\right )
\ee
are introduced. The solutions to equations (28), keeping in mind that
$\gm_0\ll\om_0$ and $\dlt_L\ll\om_0$, are
$$
S^-(t) = S^-(0)\exp\left\{ -i(\om_0 -\dlt_L) t -\gm_0 t\right \} \; ,
$$
$$
S^z(t) = -\; \frac{1}{2} + \left [ \frac{1}{2} + S^z(0)\right  ] \;
\exp(-2\gm_0 t) \; .
$$

Thus, the self-action of a radiating atom leads to the appearance of the
attenuation in the dynamics of the pseudospin operators and to the Lamb 
frequency shift. The latter can always be included in the definition of 
the transition frequency $\om_0$. Taking into consideration the attenuation, 
one usually generalizes the equations of motion by including $\gm_2$, 
instead of $\gm_0$, for $S_i^-$ and inserting $\gm_1$, instead of $2\gm_0$, 
for $S_i^z$.

\section{Stochastic mean-field quantization}

Expressing the vector potential (16) through the pseudospin variables,
we obtain the pseudospin equations (15) involving only the pseudospin
degrees of freedom. However, these equations are not yet closed, containing
the products of the pseudospin operators. If we employ the semiclassical
approximation, this would eliminate quantum fluctuations, which would make 
it impossible to describe the initial stage of radiation, when coherence
is not yet developed. In order to simplify the equations by means of a kind
of a mean-field approximation, at the same time not losing the influence
of quantum fluctuations, we use the stochastic mean-field quantization
\cite{Yukalov_8}.

We notice that in equations (15) there are the terms with different properties.
The combination 
\be
\label{31}
\xi = 2k_0\bd \cdot (\bA_{vac} + \bA_{dip} + \bA_{mat} )
\ee
describes short-range fast fluctuations. While the remaining terms containing
the pseudospin variables in the radiation vector potential (18) are of 
long-range nature. Thus, it is admissible to distinguish two types of the 
variables. One of them, $\xi$ can be treated as a stochastic variable, while
the remaining set 
\be
\label{32}       
\widetilde S \equiv \{ \bS_j : \; j = 1,2,\ldots, N \}
\ee
is treated as a collection of usual spin operators. Thus all quantities in 
equations (15) are functions of two variables, which can be denoted as
$f(\tilde{S}, \xi)$.  

Having two types of the variables, it is natural to introduce two different
averaging procedures. One is the spin averaging
\be
\label{33}
 \lgl f(\widetilde S,\xi ) \rgl \equiv {\rm Tr} \hat\rho f(\widetilde S,\xi ) \; ,
\ee
with a statistical operator $\hat{\rho}$ and the trace over the spin degrees 
of freedom. And the other is the stochastic averaging
\be
\label{34}
\lgl  \lgl f(\widetilde S,\xi ) \rgl  \rgl \equiv \int  f(\widetilde S,\xi )\;
 \cD\xi \;,
\ee
with the functional integration over the stochastic variable $\xi$.                       
           
Since the vector potential (18) characterizes long-range interactions, 
decaying as $1/r$, it is possible to use the mean-field decoupling
with respect to the spin averaging:
\be
\label{35}
 \lgl S_i^\al S_j^\bt \rgl = \lgl S_i^\al \rgl \lgl S_j^\br \rgl 
\qquad ( i \neq j) \;  ,
\ee
where the stochastic variable $\xi$ is kept untouched.  
 
Accomplishing the spin averaging of equations (15), we define the 
{\it transition function}
\be
\label{36}
 u(\br_j,t) \equiv 2 \lgl S_j^-(t) \rgl \;  ,
\ee
{\it coherence intensity}
\be
\label{37}
 w(\br_j,t) \equiv \frac{2}{N} \sum_{i(\neq j)}^N \lgl \; S_i^+(t) 
S_j^-(t) + S_j^+(t) S_i^-(t) \; \rgl \; ,
\ee
and the population imbalance
\be
\label{38}
s(\br_j,t) \equiv 2 \lgl S_j^z(t) \rgl \;   .
\ee

To simplify the presentation of the resulting equations, we introduce  
the effective external force, due to the external field,
\be
\label{39}
f_0(\br,t) \equiv - 2i \bd \cdot\bE_0(\br,t)
\ee
and the effective radiation force 
\be
\label{40}
 f_{rad}(\br,t) \equiv 2k_0 \lgl \bd \cdot\bA(\br,t) \rgl \; ,
\ee
caused by atomic interactions through the common radiation field. Passing
from the summation over atoms to the spatial integration by means of the 
replacement
$$
 \sum_{j=1}^N ~ \longrightarrow ~ \rho \int d\br \qquad \left (
 \rho \equiv \frac{N}{V} \right ) \;  ,
$$
with the integration over the whole atomic system, we get the effective 
radiation force
\be
\label{41}
  f_{rad}(\br,t) = - i\gm_0 \rho \int \left [ \;
G(\br-\br',t) u(\br',t) - 
\bbe^2_d G^*(\br-\br',t) u^*(\br',t) \; \right ] \; d\br' \;  ,
\ee
where ${\bf e}_d \equiv {\bf d}/|{\bf d}|$ and the transfer function is 
$$
G(\br,t) \equiv \Theta(ct-r) \; \frac{\exp(ik_0r)}{k_0r} \;   .
$$
 The total effective force, acting on the spin variables, is the sum
\be
\label{42}
f(\br,t) \equiv f_0(\br,t) + f_{rad}(\br,t) +\xi(\br,t) \;   .
\ee

Finally, we come to the equations for the transition function,
\be
\label{43}
\frac{\prt u}{\prt  t} = -  \left ( i\om_0 + \gm_2 \right ) u + fs \; ,
\ee
for the coherence intensity,
\be
\label{44}
 \frac{\prt w}{\prt  t} = - 2\gm_2 w + \left ( u^* f + f^* u \right ) s \;  ,
\ee
and for the population imbalance,
\be
\label{45}
 \frac{\prt s}{\prt  t} = - \; \frac{1}{2} \left ( u^* f + f^* u \right ) 
-\gm_1 ( s - \zeta )  \;  ,
\ee
where $\zeta$ is an equilibrium population imbalance for a single atom. 
As usual, these equations are assumed to be complimented by the corresponding 
initial conditions and, if necessary, boundary conditions.

\section{Scale separation approach}

The following analysis of equations (45) can be done by using the scale 
separation approach \cite{Yukalov_7,Yukalov_8,Yukalov_9,Yukalov_10}
that is a generalization to stochastic differential equations with multiple 
scales of the Krylov-Bogolubov averaging method \cite{Bogolubov_11}. Partial
differential equations can also be treated by this method 
\cite{Yukalov_8,Yukalov_12}. Here we delineate the idea of the scale separation 
approach, keeping in mind equations (45). 

The separation of scales is based on the existence of small parameters. Thus,
the influence of the external field is small according to equation (22). The
attenuations are also assumed to be small, so that
\be
\label{46}
 \frac{\gm_0}{\om_0} \ll 1 \; , \qquad \frac{\gm_1}{\om_0} \ll 1 \; , 
\qquad \frac{\gm_2}{\om_0} \ll 1 \;  .
\ee
Equations (45) can be written in the form
\be
\label{47}
 \frac{\prt u}{\prt  t} = f_u \; ,  \qquad   \frac{\prt w}{\prt  t} = f_w \; , 
\qquad    \frac{\prt s}{\prt  t} = f_s \;  ,
\ee
in which
$$
 f_\al = f_\al(u,w,s,\xi,t) \qquad (\al = u,w,s ) \;  ,
$$
The right-hand sides here are such that, if all small parameters tend to zero, 
then
\be
\label{48}
 f_w \ra 0 \; , \qquad f_s \ra 0 \;  ,
\ee
while $f_u$ remains finite. This means that the functions $w$ and $s$ are
quasi-integrals of motion, or slow variables, while $u$ is a fast variable.
The equation for $u$ is solved, with the slow variables $w$ and $s$ kept
fixed, which defines $u = u(w,s,\xi,t)$. Then this solution is substituted 
into the equations for the slow variables, with averaging their right-hand 
sides over the fast variable and over the stochastic variable, according 
to the rule
\be
\label{49}
 \overline f_\al(w,s) \equiv \lim_{\tau\ra\infty} \; 
\frac{1}{\tau} \int_0^\tau \lgl \lgl\; f_\al(u,w,s,\xi,t)\; \rgl\rgl\; dt \;  .
\ee
This results in the guiding-center equations for the slow variables:
\be
\label{50}
  \frac{\prt w}{\prt  t} = \overline f_w \; ,  \qquad   
\frac{\prt s}{\prt  t} = \overline f_s \;  .
\ee
If necessary, it is possible to find corrections to the guiding centers
\cite{Yukalov_8,Yukalov_9,Yukalov_10}.

\section{Magnetic dipole transitions}

In the previous sections, it has been assumed that the resonant atoms 
experience electric dipole transitions. Now we consider magnetic dipole 
transitions. We show that, despite their difference, both these transitions
lead to the same kind of evolution equations. This consideration is also 
of importance for answering whether spin systems could demonstrate the 
occurrence of superradiance in the same way as atoms. It turns out that 
there is a principal difference between atomic and spin systems, since the 
latter, in addition to the interactions through the common radiation field, 
possess rather strong direct dipole spin interactions destroying coherence
\cite{Yukalov_13,Yukalov_14,Yukalov_15,Yukalov_16}. 
 
In the case of magnetic dipole transitions, the atom-field interaction
is given by the Hamiltonian
\be
\label{51}
\hat H_{af} = - \sum_{j=1}^N \bM_j \cdot \bB_j \;   ,
\ee
where ${\bf B}_j = {\bf B}({\bf r}_j,t)$, the magnetic moment can be 
written \cite{Yukalov_8,Yukalov_14} as
\be
\label{52}
 \bM_j = \vec{\mu}^* S_j^- + \vec{\mu} S_j^+ + \vec{\mu}_0 S_j^z \; ,
\ee
and the total magnetic field is the sum 
\be
\label{53}
  \bB = \bH_0 + \bH + \bH_{vac} + \bH_{mat} 
\ee
of an external field, radiation field, vacuum field, and the field of 
the matter which the atoms are inserted in. 

The Heisenberg equations of motion yield
$$
\frac{dS_j^-}{dt} = - i\om_0 S_j^- +
i \left ( \vec{\mu}_0 S_j^- - 2\vec{\mu} S_j^z \right ) \cdot \bB_j \; ,
$$
\be
\label{54}
 \frac{dS_j^z}{dt} = i
\left ( \vec{\mu} S_j^+ - \vec{\mu}^* S_j^- \right ) \cdot \bB_j \;   .
\ee

The total vector potential is the sum 
\be
\label{55}
 \bA = \bA_{rad} +  \bA_{vac} + \bA_{mat} \;   .
\ee
The first term, caused by atomic radiation, is
\be
\label{56}
 \bA_{rad}(\br,t)  = 
\frac{1}{c} \int \bj \left ( \br', t -\; \frac{|\br-\br'|}{c} \right ) \;
\frac{d\br'}{|\br-\br'|} \;  ,
\ee
with the current density
\be
\label{57}
 \bj(\br,t) = - c \sum_{j=1}^N \bM_j(t) \times \nabla \dlt(\br-\br_j) \;  .
\ee
Taking into account magnetic momentum (52) makes it possible to rewrite
the vector potential (56) in the form
\be 
\label{58}
 \bA_{rad} = \bA_+ + \bA_- + \bA_z  \;  ,
\ee
in which
\be
\label{59}
 \bA_+(\br_i,t) = -  \sum_{j(\neq i)}^N  \frac{1}{r_{ij}^2} \left ( 1 +
\frac{r_{ij}}{c} \; \frac{\prt}{\prt t} \right ) \bn_{ij} 
\times \vec{\mu} S_j^+ \left ( t - \; \frac{r_{ij}}{c} \right ) \;  ,
\ee
${\bf A}_-$ is the Hermitian conjugate to ${\bf A}_+$, 
$$
 \br_{ij} \equiv \br_i - \br_j \; , \qquad 
\bn_{ij} \equiv \frac{\br_{ij}}{ r_{ij}} \; \qquad
 r_{ij} \equiv | \br_{ij} | \; ,
$$
and the last term of equation (58) is
\be
\label{60}
 \bA_z(\br_i,t) = - \sum_{j(\neq i)}^N  
\frac{\bn_{ij}}{r_{ij}^2} \; \times 
\vec{\mu}_0 S_j^z \left ( t - \; \frac{r_{ij}}{c} \right ) \; .
\ee
 
The radiation magnetic field becomes
\be
\label{61}
 \bH_{rad} \equiv \nabla \times \bA_{rad} = \bH_+ + \bH_- + \bH_z \;  ,
\ee
with
$$
\bH_+(\br_i,t) = -  \sum_{j(\neq i)}^N \left [ \;
\frac{\vec{\mu}-(\vec{\mu}\cdot\bn_{ij})\bn_{ij}}{c^2r_{ij}} \; 
\frac{\prt^2}{\prt t^2} \; + \right.
$$
\be
\label{62}
\left. + \;
\frac{\vec{\mu}-3(\vec{\mu}\cdot\bn_{ij})\bn_{ij}}{r_{ij}^3} \;
\left ( 1 + \frac{r_{ij}}{c} \; \frac{\prt}{\prt t} \right ) \right ] 
S_j^+ \left ( t - \; \frac{r_{ij}}{c} \right )
\ee
and 
\be
\label{63}
 \bH_z(\br_i,t) = -  \sum_{j(\neq i)}^N 
\frac{\vec{\mu}_0-3(\vec{\mu}_0\cdot\bn_{ij})\bn_{ij}}{r_{ij}^3} \;
S_j^z \left ( t - \; \frac{r_{ij}}{c} \right ) \; .
\ee

Let us define the effective radiation field
\be
\label{64}
 \bH_{eff} \equiv \frac{1}{4\pi} \int \bH_{rad}\; d\Om(\bn)  
\ee
as field (61) averaged over spherical angles. In this averaging, we take 
into account the properties
$$
\frac{1}{4\pi} \int [\; \vec{\mu} - (\vec{\mu}\cdot\bn)\bn\; ]\; d\Om(\bn ) = 
\frac{2}{3} \; \vec{\mu} \; ,
$$
$$
\frac{1}{4\pi} \int [\; \vec{\mu} - 3(\vec{\mu}\cdot\bn)\bn \;]\; d\Om(\bn ) = 0 \; , 
 \qquad   
\frac{1}{4\pi} \int (\vec{\mu}\cdot\bn)\bn \; d\Om(\bn ) = 
\frac{1}{3} \; \vec{\mu}  \; .
$$
This gives
\be
\label{65}
 \frac{1}{4\pi} \int \bH_z \; d\Om(\bn ) = 0 \; .
\ee
As a result, we find
\be
\label{66}
\bH_{eff} = \bH_{rad}^+ + \bH_{rad}^- \; ,
\ee
with
\be
\label{67}
  \bH_{rad}^+(\br_i,t) =  -\; \frac{2}{3} \; \vec{\mu} \sum_{j(\neq i)}^N 
\frac{1}{c^2r_{ij}} \;
\frac{\prt^2}{\prt t^2} \; S_j^+\left ( t - \; \frac{r_{ij}}{c} \right ) \;  .
\ee
Similarly to equations (22), we assume that
\be
\label{68}
 \frac{ | \vec{\mu}_0 \cdot\bB|}{\om_0} \ll 1 \; , \qquad
 \frac{ | \vec{\mu} \cdot\bB|}{\om_0} \ll 1 \; ,
\ee
hence the Born approximation can be invoked giving, similarly to section 2, 
\be
\label{69}
 \bH_{rad}^+(\br_i,t) = \frac{2}{3} \; k_0^3  \vec{\mu}  \sum_{j(\neq i)}^N 
G^*(\br_{ij},t) S_j^+(t) \;  ,
\ee
with the transfer function $G$ introduced in section 4. 

The radiation field (61) can be represented as a sum
\be
\label{70}
\bH_{rad} \equiv \bH_{eff} + \bH_{dip}
\ee
of the effective field (66) and the remaining part, for which
$$
 \int \bH_{dip}\;  d\Om(\bn ) = 0 \;  .
$$

As in section 4, we introduce the effective forces acting on the atoms,
caused by the external field,
\be
\label{71}
 f_0(\br,t) \equiv - 2i \vec{\mu} \cdot \bH_0(\br,t) \; ,
\ee
due to radiation,
\be
\label{72}
 f_{rad}(\br,t) \equiv - 2i \lgl\; \vec{\mu} \cdot \bH_{eff}(\br,t)\; \rgl \;  ,
\ee
and corresponding to local fluctuations
\be
\label{73}
\xi \equiv -2i \vec{\mu} \cdot ( \bH_{vac} + \bH_{dip} + \bH_{mat} ) \;  .
\ee
Defining the natural linewidth
\be
\label{74}
 \gm_0 \equiv \frac{2}{3} \; | \; \vec{\mu} \; |^2 k_0^3 \; ,
\ee
we obtain the radiation force
\be
\label{75}
 f_{rad}(\br,t) = - i\gm_0 \rho \int [ \;
G(\br-\br',t) u(\br',t) + \bbe_\mu^2 G^*(\br-\br',t) u^*(\br',t) \; ] \;
d\br' \; ,
\ee
in which ${\bf e}_\mu \equiv \vec{\mu}/|\; \vec{\mu} \;|$. 

This force (75) enjoys the same form as that in equation (41) due to 
electric dipole transitions. The following analysis can be done in complete
analogy with the case of electric dipole transitions, just replacing
the electric transition dipole ${\bf d}$ by the magnetic dipole $\vec{\mu}$.

\section{Coherent and incoherent radiation}

Some time ago, there has been a discussion in literature on the role of 
external fields in their possibility to influence the intensity of radiation.
In particular, this problem has arisen in the study of gamma radiation
of M\"{o}ssbauer nuclei inside magnetic materials with large magnetic 
moments \cite{Nikolaev_17}. We shall investigate this problem in the next 
section, but meanwhile, we need to recall the general form of the radiation 
intensity to be studied later. 
 
Below, we keep in mind electric dipole transitions, since, as has been 
explained above, mathematics for both, electric as well as magnetic dipoles, 
is the same. The local radiation intensity is defined as
\be
\label{76}
 I(\bn,t) = \lgl \; : \bn \cdot \bS(\br,t) : \; \rgl r^2 \;  ,
\ee
where the columns denote normal ordering, ${\bf n} \equiv {\bf r}/r$, and
the Poynting vector is 
\be
\label{77}
  \bS \equiv \frac{c}{8\pi} \left ( \bE_{rad} \times \bH_{rad} -
\bH_{rad} \times \bE_{rad} \right ) \; .
\ee
The radiation fields are given by the expressions
\be
\label{78}
 \bE_{rad} = - \; \frac{1}{c} \; \frac{\prt\bA_{rad}}{\prt t} \; ,
\qquad \bH_{rad}  = \nabla \times \bA_{rad} \;  .
\ee
In the wave zone, one uses the approximation
\be
\label{79}
 |\br - \br_j | \simeq r - \bn \cdot \br_j \qquad 
( r \gg | \br_j | ) \;  .
\ee
Then the radiation vector potential can be written as
\be
\label{80}
\bA_{rad} = \bA_+ + \bA_- \; ,
\ee
with 
\be
\label{81}
  \bA_+(\br,t) \simeq i\; \frac{k_0}{r} \; \bd 
\sum_{j=1}^N S_j^+ \left ( t - \; \frac{r-\bn\cdot\br_j}{c} \right )
\ee
and ${\bf A}_-$ being Hermitian conjugate to ${\bf A}_+$. Using this, 
we get the electric radiation field
\be
\label{82}
\bE_{rad} \simeq -i k_0 ( \bA_+ - \bA_- )
\ee
and magnetic radiation field
\be
\label{83}
 \bH_{rad} \simeq \bn \times \bE_{rad} \;  .
\ee

The radiation intensity (76) becomes
\be
\label{84}  
 I(\bn,t) = \frac{cr^2}{4\pi} \; \lgl \; : \bE_{rad}^2 -
(\bn \cdot \bE_{rad} )^2 : \; \rgl \;  ,
\ee
which, treating the retardation in the Born approximation, is reduced to 
\be
\label{85}
 I(\bn,t) = 
2\om_0 \gm_0 \sum_{ij}^N \vp_{ij}(\bn) \lgl \; S_i^+(t) S_j^-(t) \; \rgl \;  ,
\ee
where the formfactor
\be
\label{86}
\vp_{ij}(\bn) \equiv \frac{3}{8\pi} \; | \bn \times \bbe_d |^2
\exp ( ik_0 \bn \cdot \br_{ij} )
\ee
is introduced. 

The radiation intensity averaged over random fluctuations and the period of fast oscillations
\be
\label{87}
\overline I(\bn,t) \equiv \frac{\om_0}{2\pi} \int_t^{t+2\pi/\om_0}
\lgl \lgl \; I(\bn,t') \; \rgl \rgl \; dt'
\ee
can be presented as a sum 
\be
\label{88}
\overline I(\bn,t) = I_{inc}(\bn,t) + I_{coh}(\bn,t)
\ee
of the incoherent radiation intensity
\be
\label{89}
I_{inc}(\bn,t) = \om_0 \gm_0 \sum_{j=1}^N \vp_{jj}(\bn)
[\; 1 + s_j(t) \; ]
\ee
and the coherent radiation intensity
\be
\label{90}
I_{coh}(\bn,t) = \frac{1}{2} \; \om_0 \gm_0 \sum_{i\neq j}^N \vp_{ij}(\bn)
\overline{u_i^*(t) u_j(t) } \; .
\ee
The diagonal formfactor is
\be
\label{91}
  \vp_{jj}(\bn) = \frac{3}{8\pi} \; 
| \bn \times \bbe_d |^2 \equiv \vp(\bn) \; .
\ee

The total radiation intensity, integrated over the spherical angles, 
\be
\label{92}
  I(t) \equiv \int \overline I(\bn,t) \; d\Om(\bn) =
I_{inc}(t) + I_{coh}(t) \; ,
\ee
consists of the incoherent part
\be
\label{93}
I_{inc}(t) = \om_0 \gm_0 \sum_{j=1}^N [\; 1 + s_j(t) \; ]
\ee
and the coherent part
\be
\label{94}
 I_{coh}(t) = \frac{1}{2}\; \om_0 \gm_0 \sum_{i\neq j}^N \vp_{ij}
\overline{u_i^*(t) u_j(t) } \;  .
\ee
Here the shape factor is defined as
\be
\label{95}
 \vp_{ij} \equiv \int \vp_{ij}(\bn) \; d\Om(\bn) \; , \qquad
 \vp_{jj} \equiv \int \vp(\bn) \; d\Om(\bn) = 1 \; .
\ee

In this way, we see that to calculate the radiation intensity, we need 
to find the solutions to the evolution equations (43), (44), and (45).

\section{Influence of external fields}

Now we shall study how external fields influence the radiation intensity.
We are interested in the permanent action of external fields, because 
of which we shall not pay attention to short temporary coherent effects.
Then the evolution equations (45) can be represented as
$$
\frac{\prt u}{\prt t} = - (i\om_0 + \gm_2) u - 
2i s \bd \cdot \bE_0(\br,t) \; ,
$$
$$
\frac{\prt w}{\prt t} = 
- 2\gm_2 w - 2is ( u^*\bd - u\bd^* ) \cdot \bE_0(\br,t) \; ,
$$
\be
\label{96}
 \frac{\prt s}{\prt t} = 
i ( u^*\bd - u\bd^* ) \cdot \bE_0(\br,t) - \gm_1(s-\zeta)  \;  .
\ee
The external field 
\be
\label{97}
\bE_0(\br,t) = \bE_0 + \bE_1 e^{i(\bk\cdot\br -\om t)} +
\bE_1^* e^{-i(\bk\cdot\br -\om t)}
\ee
consists of a constant and an alternating field. As usual, there are
the following small parameters
\be
\label{98}
 \frac{\gm_1}{\om_0} \ll 1 \; , \qquad   \frac{\gm_2}{\om_0} \ll 1 \; , 
\qquad   \frac{|\bd\cdot\bE_0|}{\om_0} \ll 1 \; , \qquad 
\frac{|\bd\cdot\bE_1|}{\om_0} \ll 1 \; .
\ee
The alternating field is tuned to resonance, so that the resonance condition
\be
\label{99}
 \frac{|\Dlt|}{\om_0} \ll 1 \qquad ( \Dlt \equiv \om - \om_0 )
\ee
is valid. 

The problem can be solved in the scale separation approach, as explained in
section 5. According to the existing small parameters, the variable $u$ is
treated as fast, while $w$ and $s$, as slow. For the fast variable, we get
\be
\label{100}
 u = u_0 e^{-(i\om_0+\gm_2)t} - 2is \bd \cdot \int_0^t \bE_0(\br,t-t') 
e^{-(i\om_0+\gm_2)t'} \; dt' \; ,
\ee
where $u_0$ is an initial value of $u$. In view of field (97), we find
$$
u = - \; \frac{2s\bd\cdot\bE_0}{\om_0-i\gm_2} + 
\frac{2s\bd \cdot\bE_1}{\Dlt+i\gm_2} \; e^{i(\bk\cdot\br-\om t)} +
$$
\be
\label{101}
 +
\left ( u_0 +  \frac{2s\bd\cdot\bE_0}{\om_0-i\gm_2} -
\frac{2s\bd \cdot\bE_1}{\Dlt+i\gm_2} \; e^{i\bk\cdot\br} \right ) \; 
e^{-(i\om_0+\gm_2)t} \; .
\ee
For the guiding center of the slow variable $s$, we obtain the equation
\be
\label{102}
 \frac{\prt s}{\prt t} = - \gm_1^* ( s - \zeta^* ) \;  ,
\ee
in which
\be
\label{103}
 \gm_1^* = \gm_1 + 16\gm_2 \left ( 
\frac{|\bd\cdot\bE_0|^2}{\om_0^2+\gm_2^2} +  
\frac{|\bd\cdot\bE_1|^2}{\Dlt^2+\gm_2^2} \right ) \; , \qquad 
\zeta^* = \frac{\gm_1}{\gm_1^*} \; \zeta \; .
\ee
The solution to equation (102) is
\be
\label{104}
 s = s_0 e^{-\gm_1^* t} + 
\left ( 1 - e^{-\gm_1^* t} \right ) \zeta^* \;  .
\ee
Here $s_0$ is an initial value of $s$. For simplicity, we accept a uniform
initial condition $s_0({\bf r}) \equiv s({\bf r},0) = s_0$. Then solution
(104) is also uniform, $s({\bf r},t) = s(t)$. 

Averaging over fast oscillations, we have
$$
 \overline{u_i^* u_j} = 4s_i s_j \left (   
\frac{|\bd\cdot\bE_0|^2}{\om_0^2+\gm_2^2} +  
\frac{|\bd\cdot\bE_1|^2}{\Dlt^2+\gm_2^2}\; e^{-i\bk\cdot\br_{ij}}
\right ) \;  ,
$$
which should be substituted into the radiation intensity (94). The incoherent
radiation intensity is
\be
\label{105}
 I_{inc}(\bn,t) = N \om_0 \gm_0 \vp(\bn) [\; 1 + s(t) \; ] \;  .
\ee
For the coherent radiation intensity, we find
\be
\label{106}
I_{coh}(\bn,t) = 2N^2 \om_0 \gm_0 \vp(\bn) s^2(t) 
\left [\; F(k_0\bn) \; \frac{|\bd\cdot\bE_0|^2}{\om_0^2+\gm_2^2} +
F(k_0\bn - \bk) \; \frac{|\bd\cdot\bE_1|^2}{\Dlt^2+\gm_2^2} \; \right ] \;  ,
\ee
with the formfactor
\be
\label{107}
 F(\bk) \equiv \left |\; \frac{1}{N} \sum_{j=1}^N e^{i\bk\cdot\br_j} \;
\right |^2 \;  .
\ee

In the case of a cylindrical sample of radius $R$ and length $L$, with
$\bk=k_0\bbe_z$, the formfactors in equation (106) read
$$
F(k_0\bn) = \frac{\lbd^4}{\pi^4 R^2 L^2 \sin^2\vth \cos^2\vth} \;
J_1^2 \left ( \frac{2\pi R}{\lbd} \; \sin\vth \right ) \sin^2 \left (
\frac{\pi L}{\lbd} \; \cos\vth \right ) \;,
$$
$$
F(k_0\bn - \bk) = \frac{\lbd^4}{\pi^4 R^2 L^2 \sin^2\vth (1-\cos\vth)^2} \;
J_1^2 \left ( \frac{2\pi R}{\lbd} \; \sin\vth \right )
\sin^2 \left [ \frac{\pi L}{\lbd} \; (1 - \cos\vth ) \right ] \; ,
$$
where $\bk=k_0\bbe_z$, $k_0=2\pi/\lbd$, and $\vth$ is the angle between
$\bn$ and the axis $z$.
    
The maximal value of the formfactor is one. For the formfactor
$F(k_0 {\bf n})$, this happens under the condition
\be
\label{108}
 k_0 \bn \cdot \br_j = 2\pi n_j \qquad (n_j = 0, \pm 1, \pm 2,
\ldots ) \;  .
\ee
For instance, if $n_j = 0$, condition (108) is valid for a chain of atoms
with ${\bf r}_j$ perpendicular to the direction of ${\bf n}$.

The factor $F(k_0{\bf n} - {\bf k})$ is maximal, reaching one, when
\be
\label{109}
( k_0 \bn - \bk) \cdot \br_j = 2\pi n_j \qquad (n_j = 0, \pm 1, \pm 2,
\ldots ) \;  .   
\ee
In the present case, we consider radiating atoms. But similar conditions
take place for the case of an atomic system scattering external radiation. 
Thus, the case $n_j = 0$ corresponds to forward or backward scattering. 
When $n_j \neq 0$, the atoms have to form an ideal lattice, and equation (109)
is the condition of the Bragg scattering. The increase of the scattering
intensity under a lattice arrangement of atoms is called the Borrmann effect
\cite{Borrmann_18,Borrmann_19}. 

The total radiation intensity (92) is the sum of the incoherent intensity
\be
\label{110}
I_{inc}(t) = N \om_0 \gm_0 [\; 1 + s(t) \; ]
\ee
and the coherent radiation intensity
\be
\label{111}
I_{coh}(t) = 2N^2 \om_0 \gm_0 s^2(t) \left ( 
\frac{\vp_0|\bd\cdot\bE_0|^2}{\om_0^2+\gm_2^2} +
\frac{\vp_1|\bd\cdot\bE_1|^2}{\Dlt^2+\gm_2^2} \right ) \; .
\ee
In the latter, the notation for the shape factors
$$
\vp_0 \equiv \int \vp(\bn) F(k_0\bn) \; d\Om(\bn) \; ,
$$
\be
\label{112}
 \vp_1 \equiv \int \vp(\bn) F(k_0\bn - \bk) \; d\Om(\bn) 
\ee
are used. 

The values of the shape factors essentially depend on the shape of the
radiating sample and on the type of the dipole transition characterized
by the change of the quantum number $\Dlt m$. For instance
\begin{eqnarray}
\nonumber
\bbe_d = \left \{ \begin {array}{ll}
\bbe_z ~ & ~ (\Dlt m = 0 ) \\
\frac{1}{\sqrt{2}}\; (\bbe_x \mp i \bbe_y) ~ & ~ (\Dlt m= \pm 1 ) 
\end{array} . \right. 
\end{eqnarray}
Therefore,
\begin{eqnarray}
\nonumber
| \bn \times\bbe_d|^2 = \left \{ \begin {array}{ll}
1 - \cos^2\vth ~ & ~ (\Dlt m = 0 ) \\
1 -\frac{1}{2}\; \sin^2\vth ~ & ~ (\Dlt m= \pm 1 ) 
\end{array} . \right. 
\end{eqnarray}
In particular, for a pencil-like sample or a disk-like sample and
$\Dlt m = \pm 1$, one has \cite{Allen_1}, respectively,
\begin{eqnarray}
\label{113}
\vp_1 \simeq \left \{ \begin{array}{cc}
\frac{3\lbd}{8L} ~ & 
\left ( \frac{\lbd}{2\pi L} \ll 1 \; , ~ \frac{R}{L} \ll 1 \right ) \\
\\
\frac{3}{8} \left ( \frac{\lbd}{\pi R} \right )^2 ~ & 
\left ( \frac{\lbd}{2\pi R} \ll 1 \; , ~ \frac{L}{R} \ll 1 \right ) \; .
\end{array} \right.
\end{eqnarray}

As follows from these results, external fields increase the longitudinal
attenuation, which leads to the accelerated relaxation of $s$. They also
induce coherent radiation, but weakly influence incoherent radiation.

The alternating resonant external field usually produces a stronger effect
than the constant field. The latter can play a more important role than
the alternating field, if 
\be
\label{114}
 \vp_1 \; \frac{|\bd\cdot\bE_1|}{\gm_2} \; <  \; \vp_0 \;
\frac{|\bd\cdot\bE_0|}{\om_0} \; \ll \; 1 \; .
\ee
In other words, when
\be
\label{115}
 \frac{\vp_0\gm_2|\; \bE_0 \;|}{\vp_1\om_0|\; \bE_1 \; |} > 1 \; .
\ee

The problem of possible influence of a constant external field on the 
coherent radiation intensity was discussed 
\cite{Yukalov_7,Yukalov_8,Yukalov_20,Yukalov_21} with respect to the 
occurrence of the so-called M\"{o}ssbauer magnetic anomaly. The latter 
consists in the increase of the spectrum area, when paramagnetic state 
changes to ferromagnetic state \cite{Nikolaev_17,Babikova_22}. In some 
papers, such an increase was associated with the influence of the external 
field on the phonon characteristics of the material. This interpretation, 
however, was shown to be incorrect \cite{Yukalov_8}. 

In the case of the M\"{o}ssbauer radiation by $^{57}$Fe, the typical,
parameters are $\omega_0 \sim 10^{19}$ 1/s, $\gamma_2 \sim 10^7$ 1/s, 
$\lambda \sim 10^{-8}$ cm, $H_0 \sim 10^5$ G, and $H_1 \sim 10^{-5}$ G.
This gives
$$
 \frac{\gm_2|\; \bH_0\;|}{\om_0|\; \bH_1\; |} \sim 10^{-2} \; .
$$
If $\vp_0$ and $\vp_1$ are of the same order of magnitude, then
the appearing constant magnetic field should not play role in
such typical conditions. It could play role for a very week alternating
field of order $H_1 \sim 10^{-7}$ G. The magnetic anomaly in ferromagnets
could be due to the inhomogeneous broadening effect \cite{Yukalov_8}.

\section{Triggering dipolar waves}

If an atomic system is prepared in an excited state, then it starts radiating
by spontaneous emission that leads to the arising correlations between atoms
through the radiated field. The semiclassical approximation cannot describe 
this process. This approximation requires that an initial coherence be imposed 
upon the system, hence it is applicable only to the coherent stage of radiation.  

In the frame of the pseudospin representation, the process of radiation starts 
with spontaneous atomic radiation, resulting in the self-action, and the 
appearing atomic correlations are associated with triggering dipolar waves
\cite{Yukalov_23}. This initial stage of the atomic state can be described
by the pseudospin equations
$$
 \frac{dS_j^-}{dt} = - i\om_0 S_j^- - iS_j^z \sum_{i(\neq j)}^N 
\left ( b_{ji}S_i^+ - c_{ji} S_i^- \right ) \; ,
$$
\be
\label{116}
 \frac{dS_j^z}{dt} =  \frac{i}{2}  \sum_{i(\neq j)}^N 
\left [ \; S_j^+ \left ( b_{ji}S_i^+ - c_{ji} S_i^- \right ) -
 S_j^- \left ( b_{ji}^*S_i^- - c_{ji}^* S_i^+ \right ) \; \right ] \; ,
\ee
in which
$$
b_{ij} \equiv \frac{k_0^2}{2\pi} 
\sum_{\al\bt} d^\al D_{ij}^{\al\bt} d^\bt \; ,
\qquad
c_{ij} \equiv \frac{k_0^2}{2\pi} 
\sum_{\al\bt} d^\al \left ( D_{ij}^{\al\bt} d^\bt \right )^* \; ,
$$
$$
D_{ij}^{\al\bt} \equiv \int \Theta(ct-|\br_i-\br| ) \;
\frac{D_{\al\bt}(\br-\br_j)}{|\br_j-\br|} \;
\exp ( - ik_0 | \br_i - \br|) \; d\br \; .
$$

Triggering dipolar waves correspond to small deviations from the average 
pseudospin values:
\be
\label{117}
 S_j^\al = \lgl S_j^\al \rgl + \dlt S_j^\al \;  .
\ee
To zero order, we have the equations
\be
\label{118}
 \frac{d}{dt} \; \lgl S_j^- \rgl = - i\om_0 \lgl S_j^- \rgl \;  , \qquad
 \frac{d}{dt} \; \lgl S_j^z \rgl = 0 \; ,
\ee
whose solutions are
\be
\label{119}
   \lgl \; S_j^-(t)\; \rgl =   \lgl\; S_j^-(0) \; \rgl e^{-i\om_0 t} \; , \qquad
\lgl \; S_j^z(t) \; \rgl =   \lgl \; S_j^z(0) \; \rgl \;  .
\ee
The equations for small deviations are 
$$
\frac{d}{dt}\; \dlt S_j^- = - i\om_0 \dlt S_j^- - i \lgl S_j^z \rgl \sum_{i(\neq j)}^N 
\left ( b_{ji}\dlt S_i^+ - c_{ji} \dlt S_i^- \right ) \; ,
$$
\be
\label{120}
 \frac{d}{dt}\; \dlt S_j^z =  \frac{i}{2}  \sum_{i(\neq j)}^N 
\left [ \; \left ( b_{ji} \dlt S_i^+ - c_{ji} \dlt S_i^- \right ) \lgl  S_j^+ \rgl -
 \left ( b_{ji}^* \dlt S_i^- - c_{ji}^* \dlt S_i^+ 
\right ) \lgl S_j^- \rgl \; \right ] \;  .
\ee

Substituting into equations (120) the Fourier transforms for the deviations,
\be
\label{121}
\dlt S_j^- = \sum_k \dlt S_k^- e^{i\bk\cdot\br_j} \; , \qquad 
\dlt S_j^+ = \sum_k \dlt S_k^+ e^{-i\bk\cdot\br_j} \; ,
\ee
and for the coefficients
$$
b_{ij} = \frac{1}{N} \sum_k b_k e^{i\bk\cdot\br_{ij}} \; , \qquad 
c_{ij} = \frac{1}{N} \sum_k c_k e^{i\bk\cdot\br_{ij}} \;    ,
$$
we obtain the equations
$$
\frac{d}{dt}\; \dlt S_k^- = - i\mu_k \dlt S_k^- - i\lbd_k \dlt S_k^+ \; ,
$$
\be
\label{122}
\frac{d}{dt}\; \dlt S_k^+ = i\mu_k^* \dlt S_k^+ + i\lbd_k^* \dlt S_k^- \; ,
\ee
in which
$$
\mu_k \equiv \om_0 - c_k \lgl S_j^z \rgl \; , \qquad
\lbd_k \equiv b_{-k} \lgl S_j^z \rgl \; .
$$
The solutions to equations (122) can be represented in the form
\be
\label{123}
 \dlt S_k^- = u_k e^{-i\om_k t} +  v^*_k e^{i\om_k t} \; , \qquad
\dlt S_k^+ = u_k^* e^{i\om_k t} +  v_k e^{-i\om_k t} \; ,
\ee
with the dipolar wave spectrum
\be
\label{124}
\om_k = \sqrt{ |\mu_k|^2 - |\lbd_k|^2 } \;   .
\ee
Because of the inequalities
$$
\frac{|b_k|}{\om_0} \ll 1 \; , \qquad  \frac{|c_k|}{\om_0} \ll 1 \; , \qquad  
\left | \frac{\lbd_k}{\mu_k} \right | \ll 1 \; ,
$$
the spectrum is positive, so that the dipolar waves are stable. In the 
long-wave limit, the spectrum reads
\be
\label{125}
 \om_k \simeq \om_0 + \frac{1}{2} \; \lgl S_i^z \rgl \sum_{j(\neq i)}^N
( \Re c_{ij}) (\bk\cdot\br_{ij} )^2 \;  .
\ee
The dipolar waves trigger the process of self-organization in a radiating 
atomic system.

\section{Transverse mode expansion}

When the radiation wavelength $\lambda$ is much shorter than the 
characteristic sizes of the atomic sample, the radiating beam cannot
be uniform, but separates into filaments \cite{Yukalov_23}. The details
of such a filamentation will be treated in a following section. Meanwhile,
we just accept the possible existence of such filaments and describe the 
general way of treating them. We consider the situation, when the radiation
propagates along the axis $z$ as a plane wave with the seed frequency 
$\omega = ck$.  

Let the sample be of cylindrical shape, with radius $R$ and length $L$
which are much larger than the radiation wavelength, 
\be
\label{126}
 \frac{\lbd}{R}  \ll 1 \; , \qquad  \frac{\lbd}{L}  \ll 1 \; .
\ee
Suppose that there are $N_f$ filamentary modes in the sample. Each filament 
can be surrounded by an enveloping cylinder of volume $V_f = \pi R_f^2 L$. 
The separation of the radiating beam into filaments implies that the solutions
to the evolution equations (43) to (45) can be represented as the expansions
$$
u(\br,t) = \sum_{n=1}^{N_f} u_n(r_\perp,t) e^{ikz} \; , \qquad
w(\br,t) = \sum_{n=1}^{N_f} w_n(r_\perp,t)  \; , 
$$
\be
\label{127}
 s(\br,t) = \sum_{n=1}^{N_f} s_n(r_\perp,t)  \; ,   
\ee
over the transverse modes, where $r_\perp \equiv \sqrt{x^2 + y^2}$ is the 
transverse radial variable. For each filament, one can define the averaged
solutions
$$
u(t) \equiv \frac{1}{V_f} \int u_n(r_\perp,t)\; d\br = 
 \frac{2}{R^2_f} \int_0^{R_f} u_n(r,t)\; rdr \; , 
$$
$$
w(t) \equiv \frac{1}{V_f} \int w_n(r_\perp,t)\; d\br = 
 \frac{2}{R^2_f} \int_0^{R_f} w_n(r,t)\; rdr \; ,
$$
\be
\label{128}
s(t) \equiv \frac{1}{V_f} \int s_n(r_\perp,t)\; d\br = 
 \frac{2}{R^2_f} \int_0^{R_f} s_n(r,t)\; rdr \; ,   
\ee
averaged over the related enveloping volumes.   
      
We introduce the effective coupling functions
\be
\label{129}
\al(t) \equiv \gm_0 \rho \int \Theta(ct-r) \; 
\frac{\sin(k_0r-kz)}{k_0r} \; d\br\; ,
\ee
and
\be
\label{130}
\bt(t) \equiv \gm_0 \rho \int \Theta(ct-r) \; 
\frac{\cos(k_0r-kz)}{k_0r} \; d\br \;  .
\ee
Also, we define the average stochastic variable
\be
\label{131}
\xi(t) \equiv \frac{1}{V_f} \int \xi(\br,t) e^{-ikz} \; d\br \;   .
\ee

Employing the scale separation approach, we meet the nonresonant terms
of the type
$$
(\al+ i\bt) s u^*\bbe_d^2 \; ,
$$
$$
s(\al+ i\bt) \left (u^*\bbe_d\right )^2 +
s(\al-i\bt) \left (\bbe_d^* u \right )^2 \; ,
$$
$$
\frac{1}{2} \; (\al+ i\bt) \left (u^*\bbe_d\right )^2 +
\frac{1}{2} \; (\al- i\bt) \left (\bbe_d^* u \right )^2 \; ,
$$
which give the contribution of order $\gamma_0/\omega_0$, as compared 
to resonant terms, because of which such terms can be safely neglected.

For the averaged solutions (128), we obtain the equations 
$$
\frac{du}{dt} = - i (\om_0 + \bt s ) u - (\gm_2 - \al s) u + \xi s \; , 
$$
$$
\frac{dw}{dt} = - 2 (\gm_2 - \al s) w + \left ( u^* \xi + \xi^* u \right ) s \; ,
$$
\be
\label{132}
 \frac{ds}{dt} = -\al w - \frac{1}{2} \; \left ( u^* \xi + \xi^* u \right ) s 
-  \gm_1 (s - \zeta) \;  .
\ee
  
We introduce the effective attenuation
\be
\label{133}
 \Gm \equiv \gm_2 - \al s \;  ,
\ee
taking into account collective processes, and the effective frequency
\be
\label{134}
  \Om \equiv \om_0 + \bt s \; ,
\ee
including the collective Lamb shift. The solution for the fast variable 
takes the form
\be
\label{135}
 u = u_0 e^{-(i\Om+\Gm)t} + s \int_0^t \xi(t-t') e^{-(i\Om+\Gm)t'} \; dt' \; .
\ee

The stochastic variable (131) is assumed to be zero centered,
\be
\label{136}
 \lgl \lgl \xi(t) \rgl\rgl = 0 \;  .
\ee
Also, we shall need the quantity
\be
\label{137}
 \gm_3 \equiv {\rm Re} \lim_{\tau\ra\infty} \; \frac{1}{\tau}
\int_0^\tau dt \; 
\int_0^t  \lgl \lgl \; \xi^*(t)\xi(t-t') \; \rgl\rgl e^{-(i\Om+\Gm)t'} \; dt' \; ,
\ee
playing the role of a dynamic attenuation caused by the stochastic variables.
Actually, it is admissible to set the correlation property
\be
\label{138}
 \lgl \lgl \; \xi^*(t)\xi(t') \; \rgl\rgl  = 2\gm_3 \dlt(t-t')
\ee
that reduces definition (137) to an identity. 

Following the scale separation approach, we substitute the fast 
variable (135) into the equations for slow variables and average the latter 
over time and over the stochastic variable $\xi$. This gives the equation 
for the coherence intensity
\be
\label{139}
\frac{dw}{dt} = -2 (\gm_2 - \al s ) w + 2\gm_3 s^2 \; ,
\ee
and for the population imbalance
\be
\label{140}
\frac{ds}{dt} = -\al w -\gm_3 s - \gm_1 ( s - \zeta) \;   .
\ee
These equations are applicable to all stages of atomic radiation.

\section{Emergence of coherence from chaos}

An excited atomic system passes through several qualitatively different 
dynamic stages, similarly to the relaxation of any statistical system from
a nonequilibrium state \cite{Yukalov_24}. The first is the interaction stage, 
\be
\label{141}
0 < t < t_{int} \qquad (interaction \; stage ) \;  ,
\ee
during which atoms begin spontaneous radiation, but radiate independently 
from each other, having yet no time for developing mutual interactions. 
The initial values of the coherence intensity and population imbalance 
practically do not change,
\be
\label{142}
  w(t_{int}) \approx w_0 \; , \qquad s(t_{int}) \approx s_0 \;  .
\ee
The interaction time is very short, being of order $t_{int} \sim a/c$, where
$a$ is the mean interatomic distance. 

After the interaction time, atoms start feeling each other through the photon 
exchange, but there is no yet correlation between them, so that they radiate
rather independently from each other. This is the chaotic quantum stage, 
lasting in the interval
\be
\label{143}
 t_{int}   < t <  t_{coh}  \qquad (chaotic  \; stage ) \;  ,
\ee
before the coherence time, when strong correlations between atoms appear. 
After the interaction time, the effective coupling functions (129) and (130) 
grow as
$$
\al(t) \ra g\gm_2 \; , \qquad \bt(t) \ra g' \gm_2 \qquad ( t > t_{int}) \; ,
$$
where the dimensionless coupling parameters are
\be
\label{144}
g \equiv \rho \; \frac{\gm_0}{\gm_2} \int \frac{\sin(k_0r-kz)}{k_0 r} \; d\br
\ee
and
\be
\label{145}
g' \equiv \rho \; \frac{\gm_0}{\gm_2} \int \frac{\cos(k_0r-kz)}{k_0 r} \; d\br \;   .
\ee
The integration here is over $V_f$. If no initial coherence is imposed
on the excited system, so that
\be
\label{146}
 w_0 \equiv w(0) = 0 \; , \qquad s_0 \neq 0 \;  ,
\ee
then the evolution equations are
\be
\label{147}    
 \frac{dw}{dt} = 2 \gm_3 s^2 \; , \qquad 
\frac{ds}{dt} = - (\gm_1 + \gm_3) s + \gm_1 \zeta \;  .
\ee
The coherence time satisfies the inequality
\be
\label{148}
 ( \gm_1 + \gm_3 ) t_{coh} \ll 1 \;  .
\ee
At this chaotic stage, the solutions to equations (147) are
\be
\label{149}
 w  \simeq 2\gm_3 s_0^2 t \; , \qquad
s \simeq s_0 - [ (\gm_1 + \gm_3) s_0 - \gm_1 \zeta ] t  \; .
\ee

The coherence time corresponds to the point where the collective term
in equation (139) becomes comparable with the chaotic term due to quantum
fluctuations, that is, when
\be
\label{150}
 \gm_2 ( gs - 1 ) w = \gm_3 s^2 \qquad ( t = t_{coh} ) \;  .
\ee
This may happen under a sufficient initial atomic excitation and a strong
coupling, when $g s_0 > 1$. Then the coherence time is  
\be
\label{151}    
 t_{coh} = \frac{s_0}{2[\gm_2(gs_0-1)s_0+\gm_3s_0+\gm_1(s_0-\zeta)]} \;  .
\ee
If the coupling parameter is large, then
\be
\label{152}
 t_{coh} \simeq \frac{T_2}{2g s_0} \qquad ( gs_0 \gg 1) \;  ,
\ee
where $T_2 \equiv \gamma_2^{-1}$. We may notice that the system can reach 
the coherence time only if it is initially excited, so that $s_0 > 0$. At 
the end of the chaotic stage, the solutions are
\be
\label{153}
 w(t_{coh}) \simeq 2 \gm_3 s_0^2 t_{coh} \; , \qquad 
s(t_{coh}) \simeq s_0 \;  .
\ee

As soon as coherence develops in the system, the coherent stage comes 
into play, exhibiting superradiance occurring in the interval
\be
\label{154}
 t_{coh} < t < T_2 \qquad (coherent \; stage) \;  .
\ee
At this stage, collective effects are dominant, so that 
$\gamma_1 \ll g \gamma_2$ and $\gamma_3 \ll g \gamma_2$. The dynamics
is described by the equations   
\be
\label{155}
 \frac{dw}{dt} = - 2\gm_2 ( 1 - gs ) w \; , \qquad
\frac{ds}{dt} = - g\gm_2 w \;  ,
\ee
whose solutions are
\be
\label{156}
 w = \left ( \frac{\gm_p}{g\gm_2} \right )^2 
{\rm sech}^2 \left ( \frac{t-t_0}{\tau_p} \right ) \; , \qquad  
s = \frac{1}{g}\; -  \; \frac{\gm_p}{g\gm_2} \; 
\tanh \left ( \frac{t-t_0}{\tau_p} \right )  \;  .
\ee
Here the notations are used:
\be
\label{157}
 \gm_p^2 = \gm_g^2 + 2g^2 \gm_2^2 \gm_3 s_0^2 t_{coh} \; , \qquad
\gm_g = (gs_0 - 1) \gm_2 \;  .
\ee
The delay time, where the coherence intensity is maximal, reads
\be
\label{158}
t_0 = t_{coh} + \frac{\tau_p}{2} \; 
\ln \left | \frac{\gm_p+\gm_g}{\gm_p-\gm_g} \right | \;  .
\ee
The superradiant pulse time is
\be
\label{159}
 \tau_p \equiv \frac{1}{\gm_p} = \frac{T_2}{gs_0-1} \left [ 1 - \;
\frac{g^2s_0^2\gm_3 t_{coh}}{(gs_0-1)^2} \right ] \;  .
\ee
If the atomic coupling is strong, such that $gs_0 \gg 1$, then
$$
 t_0 \simeq 5 t_{coh} \; , \qquad \tau_p \simeq 2 t_{coh} \;  ,
$$
and the delay time becomes
\be
\label{160}
 t_0 = t_{coh} \left ( 1 + 
\ln \left | \frac{2}{\gm_3 t_{coh}} \right | \right ) \;  .
\ee
 
Coherence dies out at the scale of $T_2$, after which the system relaxes
during the relaxation stage
\be
\label{161}
 T_2 < t < T_1 \qquad (relaxation \; stage) \;  ,
\ee
with the decaying solutions
\be
\label{162}
 w \simeq \left ( \frac{2\gm_p}{g\gm_2} \right )^2 \; 
\exp \left (-\; \frac{2t}{\tau_p} \right ) \; , \qquad 
s\simeq  \frac{\gm_2-\gm_p}{g\gm_2}  + \frac{2\gm_p}{g\gm_2} \;
\exp \left (-\; \frac{2t}{\tau_p} \right ) \; .
\ee
The longitudinal relaxation time is $T_1 = 1/ \gamma_1$.  

Finally, for large times
\be
\label{163}
 t > T_1 \qquad ( quasi-stationary \; stage ) \;  ,
\ee
the quasi-stationary stage settles down, described by the equations
\be
\label{164}
 \frac{dw}{dt} = - 2 \gm_2 ( 1 - gs ) w + 2\gm_3 s^2 \; , \qquad
 \frac{ds}{dt} = - g \gm_2 w - \gm_3 s - \gm_1 (s - \zeta) \;  .
\ee
Stationary solutions are defined by the zeroes of the right-hand sides 
of these equations, and small oscillations around the stationary solutions
characterize the remnants of radiation \cite{Yukalov_8}. 

In this way, when no initial coherence has been imposed upon the atomic 
system, its evolution passes through the following qualitatively different
temporal intervals: interaction stage, chaotic stage, coherent stage, 
relaxation stage, and quasi-stationary stage. Coherent radiation during
the coherent stage corresponds to pure superradiance. Here the term 
{\it pure} stresses that superradiance develops as a self-organized 
spontaneous process, without being forced with external fields. The 
self-organization of atomic radiation, due to mutual correlations through 
the common radiation field, is the essence of the Dicke effect 
\cite{Dicke_25}.

However, if at the initial moment of time, a coherent pulse is imposed 
upon the atomic system, then the chaotic stage can be essentially shortened,
depending on the pulse intensity. This regime is called triggered 
superradiance. If the initial coherent pulse is strong, then the chaotic stage 
may disappear at all, so that after the interaction stage, the coherent stage 
comes into play. Then the superradiant pulse time becomes
\be
\label{165}
 \tau_p = \frac{1}{\sqrt{\gm_g^2 + (g\gm_2)^2 w_0 } } \;  .
\ee
The evolution after the coherent stage is the same as in the case of pure 
superradiance. 

In all the cases, the transitions between different dynamic stages are not
absolutely sharp, but are rather gradual crossovers.

\section{Pulsing and punctuated superradiance}

If the atomic system is subject to non-resonant pumping supporting the 
condition
$$
 g\zeta \gg 1 \;  ,
$$
then there appears a series of superradiant pulses, because of which this 
regime can be named {\it pulsing superradiance} \cite{Yukalov_8}. After 
a series of superradiant bursts, the solutions tend to the stationary state
$$
w^* \simeq \frac{\gm_1\zeta}{\gm_2 g} \; , \qquad
s^* \simeq \frac{1}{g} 
\left ( 1 - \; \frac{\gm_3}{\gm_1 g \zeta} \right ) \;   ,
$$ 
which is a stable focus. The intervals between the superradiant pulses
approximately are given by the period
$$
T_{eff} = \pi \; \sqrt{ \frac{2T_1T_2}{g\zeta} }  \;  .
$$

Another superradiant regime can be organized by acting on the atomic system 
with $\pi$ pulses inverting the atomic population at the chosen instants of 
time. Then the regime of {\it punctuated superradiance} occurs 
\cite{Yukalov_24}.

\section{Turbulent photon filamentation}

In nonlinear media interacting with electromagnetic fields there appear
different spatiotemporal structures that are analogous to the structures
arising in many other complex nonequilibrium systems 
\cite{Yukalov_8,Lugiato_26,Cross_27,Arecchi_28,Arecchi_29,Akhmanov_30}. 
The most known among such electromagnetic structures are the optical filaments
which can be formed in passive nonlinear matter 
\cite{Arecchi_28,Arecchi_29,Akhmanov_30} and in active laser media 
\cite{Yukalov_8,Lugiato_26,Arecchi_29}.

The behaviour and characteristics of optical filaments, arising in laser media, 
essentially depend on the value of the Fresnel number $F\equiv R^2/\lbd L$, 
in which $R$ and $L$ are the internal radius (aperture radius) and effective 
length, respectively, of a cylindrical laser sample, and $\lbd$ is the optical 
wavelength. There are two types of optical filaments, regular and turbulent, 
corresponding to either low or high Fresnel numbers.

Here we consider the turbulent photon filamentation. The theory of the 
turbulent filamentation in laser media was advanced, first, on the basis 
of stationary models \cite{Emelyanov_31,Yukalov_32,Yukalov_33,Yukalov_34}
invoking the notion of an effective time-averaged energy. A more elaborate 
approach, based on realistic evolution equations, was developed later
\cite{Yukalov_8,Yukalov_35,Yukalov_36,Yukalov_37,Yukalov_38}.

As is mentioned above, there are two types of optical filaments, regular and 
turbulent, which is related to the value of the laser Fresnel number. The 
latter plays for optical systems the same role as the Reynolds number for 
moving fluids. When increasing the Reynolds number, a laminar fluid transforms 
into turbulent. In the similar manner, increasing the Fresnel number makes 
a regular filamentary structure turbulent. Optical turbulence implies, by 
analogy with the fluid turbulence, that the spatiotemporal dynamics is 
chaotic. This means that the radiating filaments are randomly distributed 
in space and are not correlated with each other.

In experiments, optical filaments are usually observed in the near-field 
cross-section of lasers. The typical picture, when varying the Fresnel number 
is as follows. At very small Fresnel numbers $F\ll 1$, there exists the
sole transverse central mode uniformly filling the laser medium. When the 
Fresnel number is around $F\sim 1$, the laser cavity can house several 
transverse modes seen as a regular arrangement of bright spots in the 
transverse cross-section. Each mode corresponds to a filament extended 
through the cylindrical volume. This filamentary structure is regular in 
space, forming ordered geometric arrays, such as polygons. The transverse 
structure is imposed by the cavity geometry, being prescribed by the 
empty-cavity Gauss-Laguerre modes. Such regular structures are well 
understood theoretically; their description is based on the field expansion
over the modal Gauss-Laguerre functions related to the cylindrical
geometry \cite{Lugiato_26}. For Fresnel numbers up to $F\approx 5$, the 
number of bright filaments follows the $F^2$ law as $F$ increases. The
regular filamentary structures have been observed in several lasers, such 
as CO$_2$ and Na$_2$ lasers \cite{Lugiato_26}. Similar structures also appear
in many passive nonlinear media, e.g. in Kerr medium and in active nonlinear 
media, as the photorefractive Bi$_{12}$SiO$_2$ crystal pumped by a laser 
\cite{Arecchi_28,Arecchi_29}.

As soon as the Fresnel number reaches $F\approx 10$, there occurs a
qualitative change in the features of the filamentary structure: The regular 
filaments become turbulent. This transition goes gradually, as a crossover, 
with the intermittent behaviour in the region $5<F<15$. The character of 
this change is again common for active nonlinear media 
\cite{Arecchi_28,Arecchi_29} as well as for lasers \cite{Huyet_39,Huyet_40}. 

At Fresnel numbers $F>15$, the arising filamentary structures become 
principally different from those existing at low Fresnel numbers. The spatial 
structures now have no relation to the empty-cavity modes. The modal expansion 
over the geometrically prescribed Gauss-Laguerre modes is no longer relevant 
and the boundary conditions have no importance. The laser medium houses 
a large number of parallel independent filaments exhibiting themselves as 
a set of bright spots randomly distributed in the transverse cross-section. 
The number of these random filaments is proportional to $F$, contrary to the 
case of low Fresnel numbers, when the number of filaments is proportional 
to $F^2$. The chaotic filaments, being randomly distributed is space, are 
not correlated with each other. Such a spatio-temporal chaotic behaviour 
is characteristic of hydrodynamic turbulence, because of which the similar 
phenomenon in optics is commonly called the {\it optical turbulence}. In 
contrast to the regime of low $F$, where the regularity of spatial structures 
is prescribed by the cavity geometry and boundary conditions imposing their 
symmetry constraints, the turbulent optical filamentation is strictly 
self-organized, with its organization emerging from intrinsic properties 
of the medium. Since the optical turbulence is accompanied by the formation 
of bright filaments with a high density of photons, this phenomenon can be
named \cite{Yukalov_35} the {\it turbulent photon filamentation}. This 
phenomenon is common for lasers as well as for photorefractive crystals 
\cite{Yukalov_8,Arecchi_28,Arecchi_29}.

The first observations of the turbulent filamentary structures in lasers, 
to my knowledge, were accomplished in the series of experiments
\cite{Korolev_41,Abrosimov_42,Korolev_43,Ishenko_44,Korolev_45} with the 
resonatorless superluminescent lasers on the vapours of Ne, Tl, Pb, N$_2$, 
and N$_2^+$. In these experiments, the typical characteristics were as 
follows: $\lbd\approx 5\times 10^{-5}$ cm, $R\approx 0.1-0.3$ cm, 
$L\approx 20-50$ cm, and $F\approx 10-100$. The number of filaments was 
$N_f\sim 10^2-10^3$, with the typical radius $r_f\approx 0.01$ cm.

The filamentary structures in large-aperture optical devices have been 
observed in several lasers, as reviewed in \cite{Huyet_39,Huyet_40}, and 
in photorefractive crystals \cite{Arecchi_28,Arecchi_29}. Numerical 
simulations have been accomplished \cite{Feng_46}. Experimental works 
mainly dealt with the CO$_2$ lasers 
\cite{Huyet_39,Huyet_40,Encinas_47,Encinas_48}, dye lasers \cite{Leyva_49}, 
and semiconductor lasers \cite{Hess_50,Hegarty_51}.

The turbulent nature of filamentation occurring in high Fresnel number
lasers was carefully studied in a series of experiments 
\cite{Pastor_52,Pastor_53,Perez_54,Perez_55,Encinas_56,Calderon_57,Calderon_58}             
with CO$_2$ lasers and dye lasers. Irregular temporal behaviour was
observed in local field measurements. It was found that the transverse
correlation length was rather short. Randomly distributed transverse
patterns generated in short times were observed, being shot-to-shot
nonreproducible. For intermediate Fresnel numbers $F\sim 10$,
instantaneous transverse structures were randomly distributed in space,
but after being temporally averaged, they displayed a kind of regularity
related to the geometrical boundary conditions. This type of combination
of irregular instantaneous patterns with the averaged or stationary
pattern, showing the remnant ordering, is understandable for the
intermediate regime in the crossover region $5<F<15$. Fully developed
optical turbulence is reached as the Fresnel number increases up to
$F\sim 100$.

The typical laser parameters are as follows. The pulsed CO$_2$ laser, 
with the wavelength $\lbd=1.06\times 10^{-3}$ cm and frequency
$\om=1.78\times 10^{14}$ s$^{-1}$, emits the pulses of $\tau_p\approx
0.7\times 10^{-7}$ s or $10^{-6}$ s. The aperture radius $R\approx 1$ cm, 
laser length $L = 100$ cm. The inversion and polarization decay rates
are $\gm_1 = 10^7$ s$^{-1}$ and $\gm_2 = 3\times 10^9$ s$^{-1}$. 
The CO$_2$ density is $\rho = 2\times 10^{18}$ cm$^{-3}$. The Fresnel 
number is $F\approx 10$. The characteristic filament radius is 
$r_f\approx 0.1$ cm.

The pulsed dye laser, with the wavelength $\lbd = 0.6\times 10^{-4}$ cm
and frequency $\om = 3.14\times 10^{15}$ s$^{-1}$, produces pulses of
$\tau_p\approx 0.5\times 10^{-6}$ s. The decay rates are $\gm_1=4\times
10^8$ s$^{-1}$ and $\gm_2 = 10^{12}$ s$^{-1}$. The cavity length is
$L\approx 20$ cm. By varying the aperture radius between $0.3$ cm and
$0.8$ cm, the Fresnel number can be changed by an order, between $F = 15$
and $F = 110$. The typical filaments radius is $r_f\approx 0.01$ cm.

The theory of turbulent photon filamentation has been developed in
\cite{Yukalov_8,Yukalov_35,Yukalov_36,Yukalov_37,Yukalov_38,Yukalov_59}.

Filaments are randomly distributed in the transverse cross-section
of the laser cavity, evolving in space and time independently of each
other. The characteristics of each filament essentially depend on the
value of the related coupling parameter $g$. For cylindric symmetry
the latter can be presented in the form
\be
\label{166}
g = 2\pi \rho \; \frac{\gm_0}{\gm_2} \;
\int_0^{R_f} r_\perp\; dr_\perp \; \int_{-L/2}^{L/2} \;
\frac{\sin(k_0\sqrt{r_\perp^2+z^2}-kz)}{k_0\sqrt{r_\perp^2+z^2}}\;
dz \; .
\ee
Keeping in mind the resonance condition $k_0\approx k$ and introducing
the variable $x=k(\sqrt{r_\perp^2+z^2}-z)$, we have
\be
\label{167}
g =2\pi\; \frac{\rho\gm_0}{k\gm_2}\; \int_0^{R_f} r_\perp\; dr_\perp\;
\int_{k r_\perp^2/L}^{kL} \; \frac{\sin x}{x}\; dx \; .
\ee
Since $\lbd\ll L$, the upper limit $kL$ in the integral (167) can be replaced 
by $kL\ra\infty$. This gives
\be
\label{168}
g = 2\pi\; \frac{\rho\gm_0}{k\gm_2} \; \int_0^{R_f} \left [
\frac{\pi}{2}\; - \; {\rm Si}\left ( \frac{k r_\perp^2}{L}\right )
\right ] r_\perp\; dr_\perp \; ,
\ee
with the integral sine
$$
{\rm Si}(x) \equiv \int_0^x \frac{\sin u}{u}\; du = \frac{\pi}{2}
+ \int_\infty^x \frac{\sin u}{u}\; du \; .
$$
Introducing the notation
\be
\label{169}
\vp \equiv \frac{\pi R_f^2}{\lbd L} \; ,
\ee
varying in the interval $0\leq\vp\leq\pi F$ and playing the role of an
effective Fresnel number for a given filament, we transform equation (168) 
to
\be
\label{170}
g(\vp) = \pi \; \frac{\rho\gm_0 L}{k^2\gm_2} \left [ \pi\vp -
\int_0^{2\vp} {\rm Si}(x)\; dx\right ] \; .
\ee

In the same manner, the coupling parameter (145) can be reduced to
\be
\label{171}
g'(\vp) = -\pi\; \frac{\rho\gm_0 L}{k^2\gm_2} \;
\int_0^{2\vp} {\rm Ci}(x)\; dx \; ,
\ee
with the integral cosine
$$
{\rm Ci}(x) \equiv \int_\infty^x \frac{\cos u}{u}\; du \; .
$$

Performing the integration we find
$$
g(\vp) = \pi\; \frac{\rho\gm_0 L}{k^2\gm_2}\; \left [ \pi\vp -
2\vp{\rm Si}(2\vp) + 1 - \cos(2\vp)\right ] \; ,
$$
$$
g'(\vp) = \pi\; \frac{\rho\gm_0 L}{k^2\gm_2}\; \left [ \sin(2\vp) -
2\vp{\rm Ci}(2\vp) \right ] \; .
$$
Thus, the coupling parameters are functions of the effective variable
(169), which, in turn, depends on the enveloping radius $R_f$ related
to the effective filament radius $r_f$. If we assume that the radiation 
intensity in the transverse cross-section of a filament is distributed 
by the Gaussian law and if we define the effective filament radius as 
the mean-square deviation from the filament axis, then we get the relation
$r_f = 0.55 R_f$.  

In general, the filaments of different radii can arise. However, some of 
them are more stable than others, because of which the overwhelming majority 
of the filaments will possess the radii close to a typical value. The 
distribution of filaments with respect to their radii and, hence,
the typical radius, can be found by invoking the general method of
probabilistic pattern selection \cite{Yukalov_8,Yukalov_36,Yukalov_60,Yukalov_61}.
Following this approach, we define the probability distribution
\be
\label{172}
p(\vp,t) = \frac{1}{Z(t)}\; \exp\{-X(\vp,t)\}
\ee
for a filament characterized by the variable $\vp$ at the moment of time
$t$. Here
\be
\label{173}
X(\vp,t) ={\rm Re}\; \int_0^t {\rm Tr}\hat J(\vp,t')\; dt'
\ee
is the {\it expansion exponent}, expressed through the Jacobian matrix
$\hat J$ of the evolution equations, and
$$
Z(t) = \int \exp\{ - X(\vp,t)\} \; d\vp
$$
is the normalizing factor. The expansion exponent (173) defines the 
{\it local expansion rate}
\be
\label{174}
\Lambda(\vp,t) \equiv \frac{1}{t}\; X(\vp,t) \; .
\ee
The latter can be represented as the sum of the local Lyapunov exponents.
The partial sum of only positive Lyapunov exponents defines the entropy 
production rate \cite{Evans_62}, which does not coincide with the local 
expansion rate (174).

Thus, the probability for the appearance of filaments, characterized
by the parameter $\vp$, is given by the probability distribution (172).
As is evident, the most probable is the filament with a typical $\vp$
satisfying the {\it principle of minimal expansion} 
\cite{Yukalov_8,Yukalov_36,Yukalov_60,Yukalov_61}
\be
\label{175}
\max_\vp p(\vp,t) \Longleftrightarrow \min_\vp X(\vp,t)
\Longleftrightarrow \min_\vp \Lambda(\vp,t) \; .
\ee
This general principle follows from the minimization of the pattern
information and can be employed for arbitrary dynamical systems.

The dynamics of turbulent photon filamentation is described by the general
evolution equations (164). Calculating the corresponding Jacobian matrix 
gives 
$$
{\rm Tr} \hat J(\vp,t) = -\gm_1 -\gm_3 - 2\gm_2(1-gs) \; ,
$$
with $g = g(\vp)$ and $s = s(t)$. For $t\gg T_1$, the expansion rate 
can be presented as
\be
\label{176}
\Lambda(\vp,t) \simeq  -\gm_1 -\gm_3 - 2\gm_2(1-gs^*) \; .
\ee
Defining the stationary state $s^*$, we get
$$
\Lambda(\vp,t) \simeq  -\gm_1 -\gm_3 -2\gm_2 (1 + |g\zeta|)  \qquad
(g\zeta\ll -1) \; ,
$$
$$
\Lambda(\vp,t) \simeq  -\gm_1 -\gm_3 - 2\gm_2\left ( 1- \;
\frac{\gm_1g\zeta}{\gm_1+\gm_3}\right ) \qquad (|g\zeta|\ll 1) \; ,
$$
\be
\label{177}
\Lambda(\vp,t) \simeq  -\gm_1 -\gm_3 -
\frac{2\gm_2\gm_3}{\gm_1 g\zeta} \qquad (g\zeta\gg 1) \; .
\ee

The stationary pumping parameter $\zeta$ is in the interval
$-1\leq\zeta\leq 1$, depending on the level of pumping. When there is
no stationary pumping, $\zeta = -1$. One says that the pumping is weak, 
if $-1 < \zeta < 0$, and it is strong, if $0 < \zeta < 1$. Keeping in mind 
that the coupling parameter $g$ is positive, we see that there exist two 
different cases, when the stationary pumping is weak or absent, $\zeta < 0$, 
and when it is strong, $\zeta > 0$. According to the principle of minimal 
expansion (175), the minimum of the expansion rate corresponds to the maximum 
of $g(\vp)$ if $\zeta < 0$, and to the minimum of $g(\vp)$, if $\zeta > 0$. 
The extrema of $g(\vp)$ are given by the equation
\be
\label{178}
{\rm Si}(2\vp) = \frac{\pi}{2} \; .
\ee

In the standard situation of absent or weak pumping, $\zeta < 0$, we
have to look for the absolute maximum of $g(\vp)$. Then equation (178) gives
$\vp = 0.96$. Therefore $R_f = 0.55\sqrt{\lbd L}$, and the 
{\it typical filament radius} is
\be
\label{179}
r_f = 0.3\sqrt{\lbd L} \; .
\ee
The number of filaments can be estimated as $N_f\approx R^2/R_n^2$, which
yields
\be
\label{180}
N_f \approx 3.3 F \; .
\ee
The linear dependence of the filament number on the Fresnel number is
characteristic of the turbulent photon filamentation.

Note that under strong stationary pumping $(\zeta > 0)$, when we need
to look for the minimum of $g(\vp)$, we would have $\vp = 2.45$, hence,
we would obtain $R_f = 0.88\sqrt{\lbd L}$ and $r_f = 0.5\sqrt{\lbd L}$.

Formula (179) for the typical filament can be compared with the radii
observed in experiments. Thus, in different vapour lasers
\cite{Korolev_41,Abrosimov_42,Korolev_43,Ishenko_44,Korolev_45}, one has
$r_f\approx 0.01$ cm. For the CO$_2$ laser and dye lasers, it was found
\cite{Pastor_52,Pastor_53,Perez_54,Perez_55,Encinas_56,Calderon_57,Calderon_58}
that $r_f\approx 0.1$ cm and $r_f\approx 0.01$ cm, respectively.
All these data are in good agreement with formula (179).

\section{Collective liberation of light}

There exist systems, called photonic band-gap materials possessing a 
prohibited band gap, where light cannot propagate. The spontaneous radiation 
of atoms, with a frequency inside the prohibited band gap, is strongly 
suppressed \cite{Yablonovich_63}. This means that the equation for the 
population difference of a single atom can be effectively represented as 
$$
 \frac{ds}{dt} = - \gm_1 ( s - s_0 ) \;  .
$$
A single initially excited atom remains excited, so that $s = s_0$, for all 
times $t > 0$.   

However, if the density of doped atoms is sufficiently high, coherent 
interactions may develop \cite{Yukalov_8}. Then atoms can start radiating 
even inside the prohibited band gap. This collective phenomenon for atoms 
with the atomic frequency inside the prohibited band gap is termed  
{\it collective liberation of light} 
\cite{Yukalov_8,Yukalov_64,Yukalov_65,Yukalov_66,Yukalov_67,Yukalov_68}. 

When the stationary population imbalance of a single atom is $\zeta = s_0$,
then the evolution equations for the ensemble of atoms, after the interaction 
stage, are
$$
\frac{du}{dt} = - ( i \Om + \Gm )u + s \xi \; ,
$$
$$
\frac{dw}{dt} = - 2 \Gm w + ( u^* \xi + \xi^* u )s  \; ,
$$
\be
\label{181}
 \frac{ds}{dt} = - g \gm_2 w - \; 
\frac{1}{2} \left ( u^* \xi + \xi^* u \right ) s - 
\gm_1 ( s - s_0 ) \; ,
\ee
where the effective collective width and collective frequency are
\be
\label{182}
 \Gm = \gm_2 ( 1 - gs ) \; , \qquad \Om = \om_0 + g' \gm_2 s \; .
\ee
These equations are to be understood as characterizing the radiation dynamics
in a separate coherent filament of volume $V_f = \pi R_f^2 L \approx \lambda L^2$. 

Employing again the scale separation approach, we solve the equation for the 
fast variable $u$, substitute the solution into the equations for the slow 
variables and average the latter equations over fast oscillations and random 
variables. Using notation (137), we obtain the equations for the slow variables
\be
\label{183}
 \frac{dw}{dt} = - 2 \Gm w + 2 \gm_3 s^2 \; , \qquad 
  \frac{ds}{dt} = - g \gm_2 w - \gm_3 s - \gm_1 ( s - s_0 ) \; .
\ee

The random variable, representing the matter, can be written 
\cite{Yukalov_68,Yukalov_69} in the form
\be
\label{184}
 \xi(t) = \frac{1}{\sqrt{N_0}} \sum_k \gm_k \left ( b_k e^{-i\om_k t}
+  b_k^\dgr e^{i\om_k t} \right ) \; ,
\ee
where $b_k$ are bosonic degrees of freedom describing the matter and $\omega_k$
is the related spectrum of collective excitations. For the bosonic operators, 
one has the averages
$$
\lgl\lgl \; b_k^\dgr b_p \; \rgl\rgl = \dlt_{kp} n_k \; , \qquad
n_k \equiv \lgl \lgl\; b_k^\dgr b_k \; \rgl \rgl \; ,
$$
\be
\label{185}
 \lgl\lgl \; b_k b_p^\dgr \; \rgl\rgl = \dlt_{kp} (1 + n_k) \; , \qquad
\lgl \lgl \; b_k b_p \; \rgl \rgl = 0 \;  ,
\ee
and the normalization condition
\be
\label{186}
 \sum_k n_k = N_{mat} \;  .
\ee

The matter is characterized by a frequency gap 
$\Delta_p \equiv \omega_2 - \omega_1$, inside which collective excitations 
are suppressed, so that
\begin{eqnarray}
\label{187}
\gm_k = \left \{ \begin{array}{ll}
0 , ~ & ~ \om_k \in ( \om_1,\om_2) \\
\gm , ~ & ~ \om_k \not\in ( \om_1,\om_2) \; .
\end{array} \right.
\end{eqnarray}
Then equations (137) and (184) yield the dynamic attenuation
\be
\label{188}
 \gm_3 = \frac{\Gm}{N_{mat}} \sum_k | \gm_k |^2 \left [ \;
\frac{n_k}{(\om_k-\Om)^2+\Gm^2} + 
\frac{1+n_k}{(\om_k+\Om)^2+\Gm^2} \; \right ] \;  .
\ee
 
For a narrow gap, such that
\be
\label{189}
 \frac{\Dlt_p}{\om_1} \ll 1 \qquad ( \Dlt_p \equiv \om_2 - \om_1 ) \;  ,
\ee
attenuation (188) simplifies to 
\be
\label{190}
\gm_3 \simeq \frac{4\gm^2\gm_2(1-gs)}{\Dlt_p^2+4\gm_2^2(1-gs)^2} \; .
\ee

When the coupling and initial excitation are sufficiently weak, such that
\be
\label{191}
 | g s_0 | \ll 1 \;  ,
\ee
then the stationary solutions for the slow variables are
$$
w^* \simeq \left ( \frac{\gm_1 s_0}{\gm_1+\gm_3} \right )^2 \;
\frac{\gm_3}{\gm_2}\;   \left [ \; 1 + 
\frac{\gm_1(\gm_1-\gm_3)}{(\gm_1+\gm_3)^2} \; g s_0 \; \right ] \; ,
$$
\be
\label{192}
 s^* \simeq   \frac{\gm_1 s_0}{\gm_1+\gm_3} 
\left [\; 1 - \; \frac{\gm_1\gm_3}{(\gm_1+\gm_3)^2} \; g s_0 \; \right ] 
\ee
and the dynamic attenuation (190) is
\be
\label{193}
 \gm_3 \simeq \frac{4\gm^2\gm_2}{\Dlt_p^2+4\gm_2^2}
\left [\; 1 -  \left ( 1 - \; \frac{8\gm_2^2}{\Dlt_p^2+4\gm_2^2} \right )
gs^* \; \right ] \; .
\ee

In the usual situation, when $\gamma \sim \gamma_1 < \gamma_2$, 
$\delta_p \gg \gamma_2$, we have $\gamma_3 \ll \gamma_1$. Therefore, in the 
case of weak coupling (191), the stationary excitation remains practically the 
same as at the beginning, $s^* \approx s_0$, which means that there is no 
radiation, hence one can say that the light is locked.

If the coupling is strong, but the initial excitation is weak, such that
\be
\label{194}
  | g s_0 | \gg 1 \; , \qquad g s_0 < 0 \; ,
\ee
then the stationary solutions are
\be
\label{195}
w^* \simeq \frac{\gm_3 s_0^2}{\gm_2| gs_0| } \; , \qquad 
s^* \simeq s_0 \left ( 1 - \; \frac{\gm_3}{\gm_1 | gs_0 |} \right )
\ee
and the attenuation is 
\be
\label{196}
 \gm_3 \simeq \frac{4\gm^2 \gm_2 | gs_0|}{\Dlt_p^2+4\gm_2^2(gs_0)^2} \;  .
\ee
Again, there is no radiation and the light remains locked, since $s^* \approx s_0$.  
 
When the initial excitation and atomic coupling are sufficiently strong,
so that
\be
\label{197}
 gs_0 \gg 1 \;  ,
\ee
then the stationary solutions are
\be
\label{198}
w^* \simeq \frac{\gm_1 s_0}{\gm_2 g} \; , \qquad 
s^* \simeq \frac{1}{g} \left ( 1 - \; \frac{\gm_3}{\gm_1 g s_0 } \right )
\ee
and the attenuation is
\be
\label{199}
  \gm_3 \simeq - \; \frac{4\gm^2\gm_2 gs^*}{\Dlt_p^2+ 4 \gm_2^2(gs^*)^2}
\simeq - \; \frac{4\gm^2\gm_2}{\Dlt_p^2+ 4 \gm_2^2} \; .
\ee
In this case, atoms radiate and de-excite to the low population imbalance
\be
\label{200}
 \lim_{t\ra \infty} s(t) = s^* \simeq \frac{1}{g} \ll s_0 \;  .
\ee
That is, there happens the collective liberation of light.

\section{Pseudospin atomic squeezing}

The effect of squeezing allows one to reduce the level of noise when 
measuring a required quantity. Generally, the notion of squeezing is 
introduced for two operators, say $\hat{A}$ and $\hat{B}$. The uncertainty
in measuring the observable quantity, corresponding to the operator $\hat{A}$,
is characterized by the operator dispersion or variance
\be
\label{201}
 {\rm var}(\hat A) \equiv \lgl \hat A^+ \hat A \rgl - 
| \lgl \hat A \rgl |^2 \;  .
\ee
The squeezing of $\hat{A}$ with respect to $\hat{B}$ is defined by the
{\it squeezing factor}
\be
\label{202}
 Q(\hat A, \hat B) \equiv 
\frac{2{\rm var}(\hat A)}{|\lgl[\hat A,\hat B] \rgl |} \;  .
\ee
The Heisenberg uncertainty relation
\be
\label{203}
  {\rm var}(\hat A) {\rm var}(\hat B) \geq 
\frac{1}{4} |\lgl[\hat A,\hat B] \rgl |^2
\ee
can be written as 
\be
\label{204}
 Q(\hat A, \hat B) Q(\hat B, \hat A) \geq 1 \; .
\ee

One says that $\hat{A}$ is squeezed with respect to $\hat{B}$ if
\be
\label{205}
 {\rm var}(\hat A) < \frac{1}{2} \; |\lgl[\hat A,\hat B] \rgl | \; ,
\ee
which implies that
\be
\label{206}
  Q(\hat A, \hat B) < 1 \;  .
\ee

Dealing with the pseudospin operators
\be
\label{207}
 S_N^z \equiv \sum_{j=1}^N S_j^z \; , \qquad
S_N^- \equiv \sum_{j=1}^N S_j^- \;  ,
\ee
we derive the evolution equations for the variables
$$
u \equiv \frac{2}{N} \sum_{j=1}^N \lgl  S_j^- \rgl =
\frac{2}{N} \; \lgl  S_N^- \rgl \; ,
$$
$$
w \equiv \frac{4}{N^2} \sum_{i\neq j}^N \lgl  S_i^+ S_j^- \rgl =
| u |^2 \; ,
$$
\be
\label{208}
 s \equiv \frac{2}{N} \sum_{j=1}^N \lgl  S_j^z \rgl =
\frac{2}{N} \; \lgl  S_N^z \rgl \;  .
\ee
Taking into account the properties
\be
\label{209}
 {\rm var}(S_N^z) = \frac{N}{4} \; ( 1 - s^2 ) \; , \qquad
\lgl \left [ S_N^z , \; S_N^- \right ] \rgl = 
\frac{N}{2} \; |\; u\; | \;  ,
\ee
we obtain the squeezing factor
\be
\label{210} 
 Q(S_N^z,S_N^-) = \frac{1-s^2}{\sqrt{w}} \;  ,
\ee
characterizing the squeezing of $S_N^z$ with respect to $S_N^-$. If this 
factor is less than one, this means that measuring the atomic imbalance $s$ 
can be done with a better accuracy than measuring the transition quantities, 
such as, e.g., coherence intensity $w$. 

Different regimes of atomic evolution result in a variety of types of temporal
behaviour of the squeezing factor (210). Its behaviour also depends on whether
the vacuum, describing the matter, is squeezed. The level of squeezing can be
regulated in the process of punctuated superradiance \cite{Yukalov_70}.

\section{Operator entanglement production}

The notion of entanglement is nowadays widely studied because of its role 
in quantum information processing and quantum computing 
\cite{Williams_71,Nielsen_72,Keyl_73}. It is necessary to distinguish 
entanglement from entanglement production \cite{Yukalov_74,Yukalov_75}.
Entanglement characterizes the state of a bipartite, or more generally, 
of a many-body system \cite{Williams_71,Nielsen_72,Keyl_73}, while 
entanglement production shows the amount of entanglement produced by an 
operation associated with an operator \cite{Yukalov_74,Yukalov_75}. Here
we consider entanglement production that plays an important role in both
quantum information processing and quantum measurements \cite{Yukalov_76}.

Suppose we need to find out how much entanglement is produced by an 
operator $\hat{A}$. The operator acts on a multidimensional Hilbert space
\be
\label{211}
 \cH = \bigotimes_{i=1}^N \cH_i \;  .
\ee
Such spaces are typical of many-body systems \cite{Yukalov_77,Birman_78}.   
The operator, acting on a non-entangled state, such as 
$$
\vp = \bigotimes_{i=1}^N \vp_i \qquad ( \vp_i \in \cH_i ) \;   ,
$$
can transfer it to an entangled state. 

The action of an operator has to be compared with the action of its 
non-entangling counterpart that is defined as
\be
\label{212}
 \hat A^\otimes \equiv 
\frac{\bigotimes_{i=1}^N\hat A_i}{({\rm Tr}_\cH \hat A)^{N-1}} \; ,
\ee
where 
\be
\label{213}
\hat A_i \equiv {\rm Tr}_{\cH/\cH_i} \hat A   
\ee
is a factor-operator obtained by tracing out from $\hat{A}$ all degrees
of freedom except one associated with the subspace ${\mathcal H}_i$. The 
denominator in equation (212) is chosen so that to preserve the 
normalization condition 
\be
\label{214}
 {\rm Tr}_\cH  \hat A^\otimes = {\rm Tr}_\cH  \hat A \;  .
\ee

The measure of entanglement production is defined as
\be
\label{215}
 \ep(\hat A) \equiv \log \; \frac{||\hat A||}{||\hat A^\otimes||} \;  ,
\ee
where $\Vert \hat{A} \Vert$ implies an operator norm. It is possible
to choose different definitions of the norm. One possibility would be to 
opt for the Hilbert-Schmidt norm
\be
\label{216}
 ||\hat A|| \equiv \sqrt{ {\rm Tr}_\cH (\hat A^+ \hat A) } \;   ,
\ee
whose advantage is that it does not depend on the basis used for calculating 
the trace.

Generally, the operator $\hat{A}$ can depend on some parameters, so that one
can study the dependence of measure (215) on this parameter. In particular,
this parameter can be time \cite{Yukalov_79}, hence, it is possible to consider
temporal behaviour of entanglement production, for instance in the process of
atomic collective radiation \cite{Yukalov_80}.  

Another possibility to investigate the evolutional entanglement production
is by considering the entanglement production due to the evolution operator
$$
 \hat U(t) = e^{-i\hat H t} \;  ,
$$
in which $\hat{H}$ is the system Hamiltonian. Following the general way, we 
define the partial factor operator
\be
\label{217}
\hat U_i(t) \equiv {\rm Tr}_{\cH/\cH_i} \hat U(t)
\ee
and construct the corresponding non-entangling evolution operator
\be
\label{218}
 \hat U^\otimes(t) = 
\frac{\bigotimes_{i=1}^N \hat U_i(t)}{[{\rm Tr}_\cH \hat U(t) ]^{N-1}} \;  .
\ee
The measure of the evolutional entanglement production reads
\be
\label{219}
 \ep(t) \equiv \ep(\hat U(t) ) = 
\log \; \frac{||\hat U(t)||}{||\hat U^\otimes(t)||} \;  .
\ee

For the Hilbert-Schmidt norms, we have 
$$
||\hat U(t)||^2 = \prod_{i=1}^N d_i \qquad (d_i \equiv {\rm dim}\cH_i ) \; ,
$$
$$
 ||\hat U^\otimes(t)||^2 = \frac{\prod_{i=1}^N ||\hat U_i(t)||^2_{\cH_i}}
{ | {\rm Tr}_\cH \hat U(t)|^{2(N-1)} } \; .
$$
Therefore the evolutional entanglement production is characterized by the measure
\be
\label{220}
\ep(t) = \frac{1}{2} \log \left \{ | \; {\rm Tr}_\cH \hat U(t) \; |^{2(N-1)}
\prod_{i=1}^N \frac{d_i}{|| \hat U_i(t) ||^2_{\cH_i} } \right \} \;   .
\ee
This measure shows to what extent the evolution operator produces entanglement.

\section{Conclusion}

A general approach is presented that can be used for both atomic as well 
as spin systems. The approach is based on the pseudospin representation
of evolution equations and employs the methods of stochastic quantization,
scale separation, transverse mode expansion, and probabilistic pattern 
selection. The generality of the approach is also in the possibility of 
treating all stages of evolution, which makes it possible to study the 
self-organization of coherence from initial chaotic fluctuations. The 
qualitatively different temporal intervals are: interaction stage, chaotic 
stage, coherent stage, relaxation stage, and quasi-stationary stage. The
regimes of pure superradiance, triggered superradiance, pulsing 
superradiance, and punctuated superradiance are analyzed. 

This approach has earlier been used for describing nonequilibrium 
coherent phenomena in spin systems 
\cite{Yukalov_6,Yukalov_7,Yukalov_8,Yukalov_9,Yukalov_14,Yukalov_74,Yukalov_75}.   
In the present paper, it is demonstrated that the same mathematical 
techniques are applicable to characterizing radiation processes in atomic 
systems. The approach is illustrated by describing several interesting 
effects, such as triggering dipolar waves, turbulent photon filamentation, 
collective liberation of light, pseudospin atomic squeezing, and operator
entanglement production. 

Although physical processes in spin and atomic systems are rather different
\cite{Yukalov_6,Yukalov_13,Yukalov_14,Yukalov_15,Yukalov_16}, the mathematical 
methods of investigation turns out to be the same, which is the advantage of 
the suggested approach. 

\vskip 5mm
{\bf Acknowledgments}

\vskip 2mm
As I mentioned in the Introduction, I included in this review the topics 
I discussed with the late I.V. Yevseyev, whose friendly advice I am so 
much missing. 
 
I appreciate useful discussions with and help from E.P. Yukalova.
   
Financial support from the Russian Foundation for Basic Research is 
acknowledged.


\newpage


\begin{thebibliography}{9}

\bibitem{Allen_1}
Allen L and Eberly J R  1975
{\it Optical Resonance and Two-Level Atoms}
(New York: Wiley)

\bibitem{Andreev_2}
Andreev A V, Emelyanov V I and Ilinsky Y A  1993
{\it Cooperative Effects in Optics} 
(Bristol: Institute of Physics)

\bibitem{Mandel_3}
Mandel L and Wolf E  1995
{\it Optical Coherence and Quantum Optics}
(Cambridge: Cambridge University)

\bibitem{Trifonov_4}
Trifonov E, Benedict M, Ermolaev V, Malyshev V and Sokolov I  1996
{\it Superradiance: Multiatomic Coherent Emission}
(Bristol: Institute of Physics)

\bibitem{Rubtsova_5} 
Rubtsova N N, Samartsev V V and Yevseyev I V  2011
{\it Coherent Transients in Optics}
(Cambridge: CRC)

\bibitem{Yukalov_6}
Yukalov V I 2002
in {\it Encyclopedia of Nuclear Magnetic Resonance} eds Grant D M and Harris R K 
Vol {\bf 9} 697 (Chichester: Wiley) 

\bibitem{Yukalov_7}
Yukalov V I  1991
{\it Laser Phys.} {\bf 1}, 85

\bibitem{Yukalov_8}
Yukalov V I and Yukalova E P  2000
{\it Phys. Part. Nucl.} {\bf 31} 561 

\bibitem{Yukalov_9}
Yukalov V I  1993
{\it Laser Phys.} {\bf 3} 870

\bibitem{Yukalov_10}
Yukalov V I  1998
{\it Phys. At. Nucl.} {\bf 61} 1882

\bibitem{Bogolubov_11}
Bogolubov N N and Mitropolsky Y A  1961
{\it Asymptotic Methods in the Theory of Nonlinear Oscillations}
(New York: Gordon and Breach) 

\bibitem{Yukalov_12}
Yukalov V I and Yukalova E P  1997
{\it Laser Phys.} {\bf 7} 1076

\bibitem{Yukalov_13}
Yukalov V I 2002
{\it Laser Phys.} {\bf 12} 1089 

\bibitem{Yukalov_14}
Yukalov V I and Yukalova E P  2004
{\it Phys. Part. Nucl.} {\bf 35} 348 

\bibitem{Yukalov_15}
Yukalov V I and Yukalova E P 2005
{\it Laser Phys. Lett.} {\bf 2} 302

\bibitem{Yukalov_16}
Yukalov V I 2005
{\it Laser Phys. Lett.} {\bf 2} 356 

\bibitem{Nikolaev_17}
Nikolaev V I and Rusakov V S  1985
{\it M\"{o}ssbauer Investigations of Ferrites}
(Moscow: Moscow University)

\bibitem{Borrmann_18}
Borrmann J  1941
{\it Phys. Z.} {\bf 42} 157

\bibitem{Borrmann_19}
Borrmann J  1950
{\it Phys. Z.} {\bf 127} 297

\bibitem{Yukalov_20}
Yukalov V I  1989
in {\it Physics of Transition Metals} ed Baryakhtar V G
Vol {\bf 2} 165 (Kiev: Naukova Dumka)

\bibitem{Yukalov_21}
Yukalov V I  1989
{\it Mod. Phys. Lett. B} {\bf 3} 1337

\bibitem{Babikova_22}
Babikova U F, Gruzin P Z, Spirin A N and Uspensky M N  1979
{\it Solid State Commun.} {\bf 32} 191 

\bibitem{Yukalov_23}
Yukalov V I and Yukalova E P  2010
{\it Phys. Rev. B} {\bf 81} 075308

\bibitem{Yukalov_24}
Yukalov V I  1991
{\it Phys. Rep.} {\bf 208} 395

\bibitem{Dicke_25}
Dicke R H  1954
{\it Phys. Rev.} {\bf 93} 99

\bibitem{Lugiato_26}
Lugiato L. A  1992
{\it Phys. Rep.} {\bf 219} 293

\bibitem{Cross_27}
Cross M C and Hohenberg P C  1993
{\it Rev. Mod. Phys.} {\bf 65} 851

\bibitem{Arecchi_28}
Arecchi F T  1995 
{\it Physica D} {\bf 86} 297

\bibitem{Arecchi_29}
Arecchi F T, Boccaletti S and Ramazza P L  1999
{\it Phys. Rep.} {\bf 318} 1

\bibitem{Akhmanov_30}
Akhmanov S A, Sukhorukov A P and Khokhlov R V  1967
{\it Phys. Usp.} {\bf 93} 19

\bibitem{Emelyanov_31}
Emelyanov V I and Yukalov V I  1986
{\it Opt. Spectrosc.} {\bf 60} 385 

\bibitem{Yukalov_32}
Yukalov V I  1988
{\it J. Mod. Opt.} {\bf 35} 35

\bibitem{Yukalov_33}
Yukalov V I  1990
{\it J. Mod. Opt.} {\bf 37} 1361

\bibitem{Yukalov_34}
Yukalov V I  1991
{\it Laser Phys.} {\bf 1} 81

\bibitem{Yukalov_35}
Yukalov V I  2000
{\it Phys. Lett. A} {\bf 278} 30

\bibitem{Yukalov_36}
Yukalov V I  2001
{\it Physica A} {\bf 291} 255

\bibitem{Yukalov_37}
Yukalov V I  2001
{\it Opt. Spectrosc.} {\bf 91} 515

\bibitem{Yukalov_38}
Yukalov V I  2001
{\it Proc. SPIE} {\bf 4605} 237

\bibitem{Huyet_39}
Huyet G and Tredicce J R  1996
{\it Physica D} {\bf 96} 209

\bibitem{Huyet_40}
Huyet G and Rica S  1996
{\it Physica D} {\bf 96} 215

\bibitem{Korolev_41}
Korolev F A, Abrosimov G V, Odintsov A I and Yakunin V P  1970
{\it Opt. Spectrosc.} {\bf 28} 290

\bibitem{Abrosimov_42}
Abrosimov G V  1971
{\it Opt. Spectrosc.} {\bf 31} 54

\bibitem{Korolev_43}
Korolev F A, Abrosimov G V and Odintsov A I  1972
{\it Opt. Spectrosc.} {\bf 33} 399

\bibitem{Ishenko_44}
Ishenko V I, Lisitsyn V N, Razhev A M, Rautian S G and Shalagin A M  1974
{\it J. Exp. Theor. Phys. Lett.} {\bf 19} 346

\bibitem{Korolev_45}
Korolev F A, Odintsov A I, Turkin E G and Yakunin V P  1975
{\it Quantum Electron.} {\bf 2} 413 

\bibitem{Feng_46}
Feng Y and Ueda K I  2003
{\it Opt. Express} {\bf 11} 632

\bibitem{Encinas_47}
Encinas-Sanz F, Leyva I and Guerra J M  2000
{\it Phys. Rev. Lett.} {\bf 84} 883

\bibitem{Encinas_48}
Encinas-Sanz F, Leyva I and Guerra J M  2000
{\it Phys. Rev. A} {\bf 62} 043821 

\bibitem{Leyva_49}
Leyva I and Guerra J M  2002
{\it Phys. Rev. A} {\bf 66} 023820

\bibitem{Hess_50}
Hess O, Koch S and Moloney J V  1995
{\it IEEE J. Quantum Electron.} {\bf 31} 35

\bibitem{Hegarty_51}
Hegarty S P, Huyet G, McInerney J G and Choquette K D  1999
{\it Phys. Rev. Lett.} {\bf 82} 1434

\bibitem{Pastor_52}
Pastor I and Guerra J M  1990
{\it Appl. Phys. B} {\bf 51} 342

\bibitem{Pastor_53}
Pastor I, P\'erez-Garcia V M, Encinas-Sanz F, Guerra J M and Vasquez L  1993
{\it Physica D} {\bf 66} 412

\bibitem{Perez_54}
P\'erez-Garcia V M and Guerra J M  1994
{\it Phys. Rev. A} {\bf 50} 1646

\bibitem{Perez_55}
P\'erez-Garcia V M, Pastor I and Guerra J M  1995
{\it Phys. Rev. A} {\bf 52} 2392

\bibitem{Encinas_56}
Encinas-Sanz F, Guerra J M and Pastor I  1996
{\it Opt. Lett.} {\bf 21} 1153

\bibitem{Calderon_57}
Calderon O G, Leyva I and Guerra J M  1998
{\it Appl. Phys. Lett.} {\bf 73} 557

\bibitem{Calderon_58}
Calderon O G, Leyva I and Guerra J M  1999
{\it IEEE J. Quantum Electron.} {\bf 35} 1

\bibitem{Yukalov_59}
Yukalov V I 2005
in {\it Trends in Spatiotemporal Dynamics in Lasers} 
eds Calderon O G and Guerra J M
(Kerala: Research Signpost) 193 

\bibitem{Yukalov_60}
Yukalov V I   2001
{\it Phys. Lett. A} {\bf 284} 91

\bibitem{Yukalov_61}
Yukalov V I  2003
{\it Physica A} {\bf 320} 149

\bibitem{Evans_62}
Evans D J and Searles D J  2002
{\it Adv. Phys.} {\bf 51} 1529

\bibitem{Yablonovich_63}
Yablonovich E  1987
{\it Phys. Rev. Lett.} {\bf 58} 2059

\bibitem{Yukalov_64}
Yukalov V I  1998
{\it  Laser Phys.} {\bf 8} 1182

\bibitem{Yukalov_65}
Yukalov V I  1999
{\it Opt. Spectrosc.} {\bf 87} 550

\bibitem{Yukalov_66}
Yukalov V I  2000
{\it Quantum Electron.} {\bf 30} 911

\bibitem{Yukalov_67}
Yukalov V I  2000
{\it Bull. Russ. Acad. Sci. Phys.} {\bf 64} 1511

\bibitem{Yukalov_68}
Yukalov V I  2001
{\it Eur. Phys. J. D} {\bf 13} 83

\bibitem{Yukalov_69}
Yukalov V I  2002
{\it Proc SPIE} {\bf 4706} 130 

\bibitem{Yukalov_70}
Yukalov V I and Yukalova E P  2004
{\it Phys. Rev. A} {\bf 70} 053828 

\bibitem{Williams_71}
Williams C P and Clearwater S H  1998
{\it Explorations in Quantum Computing} 
(New York: Springer)

\bibitem{Nielsen_72}
Nielsen M A and Chuang I L  2000  
{\it Quantum Computation and Quantum Information} 
(Cambridge: Cambridge University) 

\bibitem{Keyl_73}
Keyl M  2002
{\it Phys. Rep.} {\bf 369} 431

\bibitem{Yukalov_74}
Yukalov V I  2003
{\it Phys. Rev. Lett.} {\bf 90} 167905 

\bibitem{Yukalov_75}
Yukalov V I  2003
{\it Phys. Rev. A} {\bf 68} 022109

\bibitem{Yukalov_76}
Yukalov V I and Sornette D  2013
{\it Laser Phys.} {\bf 23} 105502

\bibitem{Yukalov_77}
Yukalov V I  2013
{\it Laser Phys.} {\bf 23} 062001 

\bibitem{Birman_78}
Birman J L, Nazmitdinov R G and Yukalov V I  2013
{\it Phys. Rep.} {\bf 526} 1 

\bibitem{Yukalov_79}
Yukalov V I  2003
{\it Mod. Phys. Lett. B} {\bf 17} 95  

\bibitem{Yukalov_80}
Yukalov V I  2004
{\it Laser Phys.} {\bf 14} 1403 
 
\end{thebibliography}
\end{document}